\documentclass[12pt]{iopart}

\usepackage{iopams}
\usepackage{color}
\usepackage{graphicx}
\usepackage{subfigure}
\usepackage{afterpage}
\usepackage{dcolumn}% Align table columns on decimal point
\usepackage{bm}% bold math
\usepackage{color}
\usepackage{float}

\expandafter\let\csname equation*\endcsname\relax

\expandafter\let\csname endequation*\endcsname\relax

\usepackage{amsmath}
\usepackage{bm}
\usepackage{amssymb}
\usepackage{siunitx}
\usepackage{cleveref}
\usepackage{tikz}
\usepackage{hhline}

\def\url#1{\expandafter\string\csname #1\endcsname}
\def\bomega{\bar{\omega}}

\newcounter{defcounter}
\usepackage{amsthm}

\newcommand{\revision}[1]{\textcolor{black}{#1}}

\begin{document}

\title{Domain Walls and Vector Solitons in the Coupled Nonlinear Schr\"odinger Equation}

\author{David D. J. M. Snee,$^1$ Yi-Ping Ma$^{1,*}$}
\address{$^1$Department of Mathematics, Physics and Electrical Engineering, Northumbria University, Newcastle upon Tyne, NE1 8ST, United Kingdom}
\ead{yiping.m@gmail.com}

\begin{abstract}
We outline a program to classify domain walls (DWs) and vector solitons in the 1D two-component coupled nonlinear Schr\"odinger (CNLS) equation without restricting the signs or magnitudes of any coefficients. The CNLS equation is reduced first to a complex ordinary differential equation (ODE), and then to a real ODE after imposing a restriction. In the real ODE, we identify four possible equilibria including ZZ, ZN, NZ, and NN, with Z (N) denoting a zero (nonzero) value in a component, and analyze their spatial stability. We identify two types of DWs including asymmetric DWs between ZZ and NN and symmetric DWs between ZN and NZ. We identify three codimension-1 mechanisms for generating vector solitons in the real ODE including heteroclinic cycles, local bifurcations, and exact solutions. Heteroclinic cycles are formed by assembling two DWs back-to-back and generate extended bright-bright (BB), dark-dark (DD), and dark-bright (DB) solitons. Local bifurcations include the Turing (Hamiltonian-Hopf) bifurcation that generates Turing solitons with oscillatory tails and the pitchfork bifurcation that generates DB, bright-antidark, DD, and dark-antidark solitons with monotonic tails. Exact solutions include scalar bright and dark solitons with vector amplitudes. Any codimension-1 real vector soliton can be numerically continued into a codimension-0 family. Complex vector solitons have two more parameters: a dark or antidark component can be numerically continued in the wavenumber, while a bright component can be multiplied by a constant phase factor. We introduce a numerical continuation method to find real and complex vector solitons and show that DWs and DB solitons in the immiscible regime can be related by varying bifurcation parameters. We show that collisions between two DB solitons with a nonzero phase difference in their bright components typically feature a mass exchange that changes the frequencies and phases
of the two bright components and the two soliton velocities.
\end{abstract}

\section{Introduction}\label{sec:intro}

The nonlinear Schr\"odinger (NLS) equation is well known to provide a universal description of the envelope dynamics of quasi-monochromatic plane waves in weakly nonlinear dispersive systems \cite{sulem2007nonlinear,ablowitz2004discrete}. The most remarkable feature of the NLS equation is complete integrability, which enables analytical solutions to be found via the inverse scattering transform for arbitrary initial conditions \cite{Shabat-Zakharov-1972,ablowitz1981solitons}. As a natural progression, coupled NLS (CNLS) equations arise when multiple waves interact nonlinearly, as already noted in early studies~\cite{manakov1974theory,zakharov1975theory}. Subsequently, such equations find prominent applications in diverse settings including birefringent optical fibers \cite{menyuk1987nonlinear}, photorefractive materials \cite{chen1997coupled,ostrovskaya1999interaction}, topological photonics \cite{ivanov2020vector}, water waves \cite{shukla2006instability,ablowitz2015interacting}, coupled sine-Gordon equations \cite{griffiths2006modulational}, and many others. Much recent interest in CNLS equations relates to Bose-Einstein condensates (BECs), where an $N$-component CNLS equation, often called the Gross-Pitaevskii equation (GPE), describes BECs with $N$ bosonic components in the mean field limit \cite{pitaevskii2016bose,kevrekidis2015defocusing}. Here we consider a 1D two-component CNLS equation with general coefficients, which is a homogeneous GPE without trapping potential in certain parameter regimes:
\begin{equation}\label{Eq:GCNLS}
\begin{aligned}
iA_{t} + d_1A_{xx} + A(g_1|A|^2 +  g_2|B|^2) = 0, \\
iB_{t} + d_2B_{xx} + B(g_3|B|^2 +  g_4|A|^2)  = 0,
\end{aligned}
\end{equation}
where $A$ and $B$ are complex fields, and $d_{1-2}$ and $g_{1-4}$ are real constants. If $d_1d_2>0$, then Eq.~(\ref{Eq:GCNLS}) is completely integrable when $d_1=d_2$, $g_1=g_4$, $g_2=g_3$ \cite{zakharov1982integrability,sahadevan1986painleve,wang2010integrable}. If in addition $g_1g_2>0$, Eq.~(\ref{Eq:GCNLS}) becomes the celebrated Manakov system \cite{manakov1974theory}, which is focusing when $d_1g_1>0$ and defocusing when $d_1g_1<0$.

A fundamental feature of nonlinear wave equations is the possible balance of dispersion and nonlinearity to produce robust localized traveling waves embedded in a single plane wave, which are known as solitons (solitary waves) \cite{zabusky1965interaction,drazin1989solitons,Ablowitz-2013}. For the scalar NLS equation, it is well known that focusing NLS admits bright solitons, while defocusing NLS admits dark (grey) solitons \cite{ablowitz2004discrete,Ablowitz-2013}. There is also a thorough catalogue of vector solitons and their interactions in the integrable CNLS equation, including bright-bright (BB) solitons in the focusing Manakov system \cite{manakov1974theory,kaup1993soliton,radhakrishnan1997inelastic,stalin2019nondegenerate} and dark-bright (DB) and dark-dark (DD) solitons in the defocusing Manakov system \cite{christodoulides1988black,afanasyev1989dynamics,kivshar1993vector,radhakrishnan1995bright,sheppard1997polarized,park2000systematic,prinari2006inverse}. Such solutions have been generalized to the symmetric CNLS equation ($d_1=d_2, \ g_1=g_3, \ g_2=g_4$) in both focusing \cite{yang1997classification,yang2000fractal,smyth2001radiative} and defocusing \cite{haelterman1994bifurcations,sheppard1997polarized} cases.

In the binary BEC setting, there is much interest in DB and DD solitons in the defocusing regime of Eq.~(\ref{Eq:GCNLS}) \cite{kevrekidis2016solitons}. Most of these studies assume $d_1=d_2$, namely equal masses for the two species, with free parameters $g_{1-4}$. The oscillation and collision dynamics of DB solitons in a harmonic trap are revealed in a pioneering theoretical study \cite{busch2001dark}. The formation of DD solitons is also predicted from the interaction of dark solitons in different components in the miscible regime \cite{ohberg2001dark}. Subsequently, families of vector solitons are produced that include not only DB and DD solitons, but also bright-antidark (BAD) and dark-antidark (DAD) solitons, where antidark means a hump rather than a dip in a nonzero background \cite{kevrekidis2004families}. DB solitons are first realized experimentally using a phase imprinting method \cite{becker2008oscillations}. Subsequently, an alternative method is proposed to generate trains of DB solitons from the counterflow of two superfluids \cite{hamner2011generation}, which motivates further theoretical and experimental studies on the interaction dynamics of DB solitons \cite{yan2011multiple,alvarez2013scattering,yan2015dark}. DB solitons are also found in Eq.~(\ref{Eq:GCNLS}) with unequal masses (dispersion coefficients) in both $d_1d_2<0$ \cite{buryak1996coupling} and $d_1d_2>0$ \cite{achilleos2014beating,charalampidis2015dark} cases. In both cases, it is found that DB solitons with a one-hump bright component could coexist with those whose bright component has two or more humps.

% Such counterflow experiments can also generate beating DD solitons, which can be related to DB solitons via $SU(2)$ rotations in the defocusing Manakov system \cite{hoefer2011dark,yan2012beating}.

In addition to various solitons reviewed above, some nonlinear partial differential equations (PDEs) like Eq.~(\ref{Eq:GCNLS}) also admit localized traveling waves connecting two different plane waves, which are known by various names including domain walls (DWs), fronts, and kinks. Such solutions play a central role in dissipative systems in various settings \cite{xin2000front,van2003front,knobloch2015spatial,malomed2021past}. They are also found in some scalar 1D conservative (dispersive) systems, possibly with dissipative (diffusive) perturbations, such as the modified Korteweg-de Vries-Burgers equation \cite{jacobs1995traveling} and the NLS equations with stimulated Raman scattering \cite{kivshar1993raman}. In general, the time evolution of an initial step between two plane waves in a dispersive system is the core topic in the field of dispersive hydrodynamics, which reformulates the original PDE as a system of hyperbolic conservation laws \cite{el2016dispersive}. In this framework, dispersive shock waves arise when the flux is convex, while rarefaction waves and DWs (kinks) arise when the flux is non-convex \cite{el2017dispersive}.

% Note that DWs (kinks) can refer to connections between two plane waves related via phase rotations, such as kinks in the sine-Gordon equation \cite{ivancevic2013sine} and phase domain walls in the 2D hyperbolic NLS equation \cite{tsitoura2018phase}, but in this paper the two plane waves connected by DWs are typically assumed nonequivalent.

In two-component 1D conservative systems, such as Eq.~(\ref{Eq:GCNLS}) in the defocusing regime, DWs can form between the two components \cite{malomed2021past}. In the optics setting, such DWs in the symmetric CNLS equation, known as polarization domain walls (PDWs), are indeed predicted theoretically \cite{haelterman1994polarization,haelterman1994vector,malomed1994optical,jovanoski2008exact}. While there are early experimental indications of PDWs \cite{pitois1998polarization,pitois1999generation}, the first direct observation of PDWs is only made recently in optical fibres with potential applications to optical data transmission \cite{gilles2017polarization}. Applications of optical bistability to the design of photonic memory devices are also explored in recent work \cite{valagiannopoulos2021angular,valagiannopoulos2022multistability}. In the binary BEC setting, DWs are found in the symmetric CNLS equation with a trapping potential \cite{kevrekidis2004families} and possibly also linear interconversion terms \cite{dror2011domain}. More generally, in the asymmetric CNLS equation, i.e., Eq.~(\ref{Eq:GCNLS}) with $d_1d_2>0$ and free parameters $g_{1-4}$, DWs are found when the immiscibility condition $g_2g_4-g_1g_3>0$ holds \cite{trippenbach2000structure,coen2001domain,alama2015domain}.

The above studies typically make two assumptions on the coefficients of Eq.~(\ref{Eq:GCNLS}) that are relevant in the optics and/or BEC setting, the first being $d_1d_2>0$, or more often $d_1=d_2$, and the second being $g_2g_4>0$, or equivalently $g_2=g_4$ by rescaling variables. However, even if both are assumed, certain parameter regimes beyond the traditional dichotomy between the focusing and the defocusing regimes are rarely explored. Moreover, in the general problem of two interacting quasi-monochromatic plane waves, $g_2g_4>0$ often applies, but $d_1d_2>0$ and $d_1d_2<0$ are both very common. As we shall see, for DWs and vector solitons, there is a simple relation between solutions found at $g_2g_4>0$ and those found at $g_2g_4<0$. Thus, in this paper, we outline a program to classify DWs and vector solitons in Eq.~(\ref{Eq:GCNLS}) without restricting the signs or magnitudes of any coefficients.

This paper is organized as follows. In Section \ref{sec:equ-sta}, we reduce Eq.~(\ref{Eq:GCNLS}) first to an ordinary differential equation (ODE) with complex coefficients, and then to an ODE with real coefficients after imposing a restriction. In the real ODE, we identify all possible equilibria and analyze their spatial stability. In Section \ref{sec:exist-dw}, we identify all possible DWs between two equilibria and construct their parameter space. In Section \ref{sec:vec-sol}, we identify three codimension-1 mechanisms for generating vector solitons in the real ODE including heteroclinic cycles, local bifurcations, and exact solutions. Heteroclinic cycles are formed by assembling two DWs back-to-back. All possible local bifurcations on the four equilibria are identified. Exact solutions include scalar bright and dark solitons with vector amplitudes. We classify these codimension-1 real vector solitons by their types and parities and discuss numerical continuation of any such solution into a codimension-0 family. We construct the parameter space of complex vector solitons depending on whether the two components are bright, dark, or antidark. In Section \ref{sec:num-con}, we introduce a numerical continuation method to find real and complex vector solitons and apply this method to DWs and DB solitons in the immiscible regime. In Section \ref{sec:pdb-collision}, we study collisions between two DB solitons with a nonzero phase difference in their bright components. The paper concludes in Section \ref{sec:conclusion}.

\section{Equilibria and their spatial stability}\label{sec:equ-sta}

To find DWs connecting two arbitrary plane waves and vector solitons embedded in an arbitrary plane wave, we apply the following traveling wave reduction to Eq.~(\ref{Eq:GCNLS}): $A=\phi(\zeta)\exp(i(k_Ax-\omega_A t))$, $B=\psi(\zeta)\exp(i(k_Bx-\omega_B t))$, where $\phi,\psi$ are complex envelopes dependent only on the comoving variable $\zeta=x-C_gt$, with $C_g \in \mathbb{R}$ the group velocity of the envelope, and $k_A,k_B\in\mathbb{R}$ and $\omega_A,\omega_B\in\mathbb{R}$ are the respective wavenumbers and frequencies. This reduces Eq.~(\ref{Eq:GCNLS}) to a coupled system of ODEs in $\zeta$:
\begin{equation}\label{Eq:ODE}
\begin{aligned}
&d_1\phi'' + i(2d_1k_A-C_g)\phi' + (\omega_A-d_1k_A^2)\phi + g_1|\phi|^2\phi +  g_2|\psi|^2\phi = 0,  \\
&d_2\psi'' + i(2d_2k_B-C_g)\psi' + (\omega_B-d_2k_B^2)\psi + g_3|\psi|^2\psi +  g_4|\phi|^2\psi = 0.
\end{aligned}
\end{equation}
In this framework, a plane wave in Eq.~(\ref{Eq:GCNLS}) corresponds to an equilibrium in Eq.~(\ref{Eq:ODE}), namely a $\zeta$-independent solution to Eq.~(\ref{Eq:ODE}). Thus, a DW connecting two plane waves in Eq.~(\ref{Eq:GCNLS}) corresponds to a heteroclinic orbit between two equilibria in Eq.~(\ref{Eq:ODE}), and a vector soliton embedded in a plane wave in Eq.~(\ref{Eq:GCNLS}) corresponds to a homoclinic orbit to an equilibrium in Eq.~(\ref{Eq:ODE}). We note that prior work, e.g.~Ref.~\cite{buryak1996coupling}, often sets $C_g=0$ since traveling waves can be obtained from stationary waves using the Galilean transformation, but we shall keep $C_g$ as a free parameter in this paper to emphasize that both DWs and vector solitons can have nonzero velocities; see e.g.~Ref.~\cite{filatrella2014domain}.

Once DWs and vector solitons are found in Eq.~(\ref{Eq:ODE}), their temporal stability in Eq.~(\ref{Eq:GCNLS}) can be determined using established methods. The temporal stability of plane waves, namely modulational instability, in Eq.~(\ref{Eq:GCNLS}) is already addressed in an early study \cite{roskes1976some}. However, a localized structure can be temporally unstable even if its plane wave background(s) are temporally stable \cite{kapitula2013spectral}. In this section, we shall focus on the general existence of DWs and vector solitons, and we shall keep such solutions even if they are temporally unstable. \revision{Although temporally unstable solutions are less interesting than temporally stable ones, they can still be important for understanding the nonlinear dynamics of modulational instability and other complex spatiotemporal dynamics. Besides, our overall strategy is to first find solutions in a (typically codimension-1) subset of the parameter space that is analytically tractable, and then use these solutions as starting points for numerical continuations into the full parameter space. Even if a starting point is temporally unstable, it should not be discarded since the numerical continuation could yield temporally stable solutions.}

\subsection{The complex ODE}

The key property of Eq.~(\ref{Eq:ODE}) that enables the existence of homoclinic orbits is its reversibility under the action
\begin{equation}\label{eq:rev-sym}
%\begin{aligned}
R_{\Theta_A,\Theta_B}(\zeta,\phi,\psi,\phi',\psi')=
(-\zeta,e^{2i\Theta_A}\phi^*,e^{2i\Theta_B}\psi^*,-e^{2i\Theta_A}\phi^{*'},-e^{2i\Theta_B}\psi^{*'}),
%\end{aligned}
\end{equation}
where $\Theta_A,\Theta_B\in\mathbb{R}$ and $*$ denotes complex conjugation. This reversibility represents a reflection symmetry in the phase space of Eq.~(\ref{Eq:ODE}) and implies the generic existence of homoclinic orbits to an equilibrium \cite{champneys1998homoclinic}. We observe that if $(\phi,\psi)$ is reversible under $R_{0,0}$, then $(e^{i\Theta_A}\phi,e^{i\Theta_B}\psi)$ is reversible under $R_{\Theta_A,\Theta_B}$. Thus, we only need to find homoclinic orbits reversible under $R_{0,0}$, and the full solution family follows from multiplying each component by a constant phase factor. We need $\Theta_A,\Theta_B\in[0,2\pi)$ to obtain all possible solutions, but we only need $\Theta_A,\Theta_B\in[0,\pi)$ to obtain all possible reversibility symmetries in Eq.~(\ref{eq:rev-sym}).

The equilibria of Eq.~(\ref{Eq:ODE}), denoted as $(\phi_0,\psi_0)\in\mathbb{C}^2$, can be classified into four types, denoted respectively as $ZZ$, $ZN$, $NZ$, and $NN$, with $Z \ (N)$ corresponding to zero (nonzero) values. There is a circle of equivalent equilibria in the complex plane for each nonzero component, so $ZZ$ represents a single equilibrium, $ZN$ or $NZ$ represents a circle of equivalent equilibria, and $NN$ represents a torus of equivalent equilibria. Thus, under suitable conditions, we expect a discrete, not continuous, family of homoclinic orbits connecting the two equilibria $(\phi_0,\psi_0)$ and $(\phi_0^*,\psi_0^*)$ that are equivalent under the reversibility $R_{0,0}$.

\revision{Our theoretical framework of seeking solutions to a nonlinear PDE via its reduction to a spatial ODE, known as spatial dynamics \cite{champneys1998homoclinic,knobloch2015spatial}, provides a systematic method to find localized structures. For generic parameter choices, neither the original PDE in Eq.~(\ref{Eq:GCNLS}) nor the spatial ODE in Eq.~(\ref{Eq:ODE}) is integrable, and this non-integrable regime is the focus of this paper. However, even in the integrable regime of Eq.~(\ref{Eq:GCNLS}), namely the scalar NLS equation or the Manakov system, spatial dynamics can still be used to reproduce scalar or vector solitons originally derived using integrable systems theory.}

\subsection{The real ODE}

Although Eq.~(\ref{Eq:ODE}) provides the general setting for finding DWs and vector solitons, it is rather technical to work with since it is a second-order ODE with two complex fields, or equivalently an 8D real dynamical system. Therefore, we will first seek DWs and vector solitons in Eq.~(\ref{Eq:ODE}) with the restriction
\begin{equation}\label{eq:restri}
C_g=2d_1k_A=2d_2k_B,
\end{equation}
and then extend these solutions to Eq.~(\ref{Eq:ODE}) without this restriction. After applying Eq.~(\ref{eq:restri}), Eq.~(\ref{Eq:ODE}) becomes the following ODE with real coefficients:
\begin{equation}\label{eq:ODE-0_R}
\begin{aligned}
&d_1\phi'' + \bomega_A\phi + g_1|\phi|^2\phi +  g_2|\psi|^2\phi = 0,\\
&d_2\psi'' + \bomega_B\psi + g_3|\psi|^2\psi +  g_4|\phi|^2\psi = 0,
\end{aligned}
\end{equation}
where
\begin{equation}\label{eq:galilean}
\bomega_A=\omega_A-\frac{C_g^2}{4d_1},\quad
\bomega_B=\omega_B-\frac{C_g^2}{4d_2}.
\end{equation}
Thus, Eq.~(\ref{eq:galilean}) implies that Eq.~(\ref{eq:ODE-0_R}) is invariant under the parameter transformation $(C_g,\omega_A,\omega_B)\rightarrow(0,\bomega_A,\bomega_B)$, which is an ODE version of the Galilean transformation. The four types of equilibria of Eq.~(\ref{eq:ODE-0_R}) are respectively \revision{given by}
\begin{equation}\label{Eq:equilibria}
\begin{aligned}
ZZ: & \ \ \phi_0=0, \, \psi_0=0;\\
ZN: & \ \ \phi_0=0, \, |\psi_0|^2=\frac{-\bomega_B}{g_3};\\
NZ: & \ \ |\phi_0|^2=\frac{-\bomega_A}{g_1}, \, \psi_0=0;\\
NN: & \ \ |\phi_0|^2=\frac{\bomega_Ag_3-\bomega_Bg_2}{g_2g_4-g_1g_3},
 \, |\psi_0|^2=\frac{\bomega_Bg_1-\bomega_Ag_4}{g_2g_4-g_1g_3}.
\end{aligned}
\end{equation}

\revision{The restriction of Eq.~(\ref{eq:ODE-0_R}) to $\phi,\psi\in\mathbb{R}$ yields the real ODE}
\begin{equation}\label{eq:ODE-0_R_real}
\begin{aligned}
&d_1\phi'' + \bomega_A\phi + g_1\phi^3 +  g_2\psi^2\phi = 0,\\
&d_2\psi'' + \bomega_B\psi + g_3\psi^3 +  g_4\phi^2\psi = 0.
\end{aligned}
\end{equation}
\revision{The equilibria of Eq.~(\ref{eq:ODE-0_R_real}) are given by Eq.~(\ref{Eq:equilibria}) but restricted to the real axis, so the $N$ component can have either sign. The possible reversibility symmetries of Eq.~(\ref{eq:ODE-0_R_real})} must be given by Eq.~(\ref{eq:rev-sym}) but restricted to real coefficients. Thus, the only possible symmetries of Eq.~(\ref{eq:ODE-0_R_real}) are up-down symmetries:
\begin{equation}\label{eq:rev-sym-real}
\begin{aligned}
&R_{s_A,s_B}(\zeta,\phi,\psi,\phi',\psi')=
(-\zeta,s_A\phi,s_B\psi,-s_A\phi',-s_B\psi'),
\end{aligned}
\end{equation}
where $s_A,s_B\in\{+,-\}$. Since $s_{A,B}=e^{2i\Theta_{A,B}}$, $s_{A,B}=+$ corresponds to $\Theta_{A,B}=0$ and $s_{A,B}=-$ corresponds to $\Theta_{A,B}=\pi/2$. Thus, an even component is already reversible under $R_{0,0}$ in Eq.~(\ref{eq:rev-sym}), while an odd component must be divided, or equivalently multiplied, by $i=e^{i\pi/2}$, to be reversible under $R_{0,0}$ in Eq.~(\ref{eq:rev-sym}).

The coefficients of Eq.~(\ref{eq:ODE-0_R_real}) can be simplified by rescaling variables. This will not be carried out explicitly, but we record two key features. We shall refer to $(d_{1-2},g_{1-4})$ as the CNLS parameters, and $(\bomega_A,\bomega_B)$, or equivalently $(C_g,\omega_A,\omega_B)$, as the traveling wave parameters. First, in Section \ref{sec:intro}, we stated that the only possible restriction on the CNLS parameters is $g_2g_4>0$ on the PDE level of Eq.~(\ref{Eq:GCNLS}). However, this restriction is optional at the ODE level of Eq.~(\ref{eq:ODE-0_R_real}) since the $g_2g_4>0$ and $g_2g_4<0$ cases are related by flipping the sign of one of the two equations in Eq.~(\ref{eq:ODE-0_R_real}). Second, $(\bomega_A,\bomega_B)$ can be multiplied by a positive constant by rescaling $(\zeta,\phi,\psi)$, so the traveling wave parameter space can be taken as a unit circle in the $(\bomega_A,\bomega_B)$-plane rather than the whole plane. Thus, $(\bomega_A,\bomega_B)$ are effectively a single parameter that can be taken as $\bomega\equiv\bomega_B/\bomega_A$ with the caveat that each $\bomega$ corresponds to two antipodal points on the unit circle.

The existence of heteroclinic and homoclinic orbits in Eq.~(\ref{eq:ODE-0_R_real}) depends crucially on the existence and stability of its equilibria, where the stability is defined purely within Eq.~(\ref{eq:ODE-0_R_real}) with respect to the spatial coordinate $\zeta$. An equilibrium must be spatially hyperbolic, namely its spatial eigenvalues are either stable or unstable, not neutrally stable, to ensure the generic existence of heteroclinic or homoclinic orbits from or to it. To obtain the spatial eigenvalues $\lambda$ of an equilibrium, we first perturb it as
\begin{equation}\label{eq:pert-equi}
\begin{bmatrix} \phi \\ \psi \end{bmatrix}=\begin{bmatrix} \phi_0 \\ \psi_0 \end{bmatrix}+\epsilon\begin{bmatrix} \phi_1 \\ \psi_1 \end{bmatrix}\mbox{e}^{\lambda\zeta}+O(\epsilon^2), \quad \epsilon \ll 1,
\end{equation}
where $[\phi_1,\psi_1]^T$ is the spatial eigenvector. Then, we substitute this into Eq.~(\ref{eq:ODE-0_R_real}) and linearize to obtain
\begin{equation}\label{eq:eig_prob}
J\begin{bmatrix} \phi_1 \\ \psi_1 \end{bmatrix}=\lambda^2I_2\revision{\begin{bmatrix} \phi_1 \\ \psi_1 \end{bmatrix}},
\end{equation}
where $I_2$ is the $2\times2$ identity matrix and
\begin{equation}\label{eq:jac_mat}
J = \left(\begin{matrix}
D_1 & -\frac{2g_2}{d_1}\phi_0\psi_0 \\
-\frac{2g_4}{d_2}\phi_0\psi_0 & D_2
\end{matrix}\right)
\end{equation}
is the Jacobian matrix with
\begin{align*}
&D_1=-\frac{1}{d_1}\left(\bomega_A+3g_1\phi_0^2+g_2\psi_0^2\right), \\
&D_2=-\frac{1}{d_2}\left(\bomega_B+g_4\phi_0^2+3g_3\psi_0^2\right).
\end{align*}
We note that Eq.~(\ref{eq:ODE-0_R_real}) as a 4D dynamical system would have a $4\times4$ Jacobian matrix, but Eqs.~(\ref{eq:eig_prob}--\ref{eq:jac_mat}) provide a simpler and equivalent formulation. The characteristic equation of the eigenvalue problem in Eq.~(\ref{eq:eig_prob}) is
\begin{equation}\label{eq:char-poly}
\frac{4g_2g_4}{d_1d_2}\phi_0^2\psi_0^2-D_1D_2+(D_1+D_2)\lambda^2-\lambda^4=0,
\end{equation}
which is a quadratic equation in $\lambda^2$. The equilibrium $(\phi_0,\psi_0)$ is spatially hyperbolic if and only if $\mathcal{R}e(\lambda) \neq 0$ for all four roots \revision{$\lambda$} of Eq.~(\ref{eq:char-poly}), \revision{or equivalently Eq.~(\ref{eq:char-poly}) has no non-positive real roots $\lambda^2$. The Jacobian matrix $J$ in Eq.~(\ref{eq:jac_mat}) is generally a $2\times2$ real matrix without any symmetry, so the configuration of its two eigenvalues $\lambda^2$ given by the two roots $\lambda^2$ of Eq.~(\ref{eq:char-poly}) is always symmetric with respect to the real axis. There can be two positive real roots, two negative real roots, one positive and one negative real roots, or two complex roots. After taking the square root,} the configuration of these four spatial eigenvalues \revision{$\lambda$} in the complex plane is always symmetric with respect to both the real and the imaginary axes, so a spatially hyperbolic equilibrium always has two stable and two unstable spatial eigenvalues; \revision{see Ref.~\cite{champneys1998homoclinic} for a detailed discussion of spatial eigenvalues in 4D reversible dynamical systems and their ramifications.}

\revision{One can similarly analyze the spatial stability in Eq.~(\ref{eq:ODE-0_R}) of any complex equilibrium in Eq.~(\ref{Eq:equilibria}). This will not be shown explicitly, but the key conclusion is that an equilibrium and its stable or unstable eigenspace always have the same phases in both $\phi$ and $\psi$. Thus, the stable or unstable manifold of a complex equilibrium in Eq.~(\ref{eq:ODE-0_R}) is equivalent to that of a real equilibrium in Eq.~(\ref{eq:ODE-0_R_real}) modulo phase rotations, so we can seek any solution tending to an equilibrium in Eq.~(\ref{eq:ODE-0_R_real}) rather than Eq.~(\ref{eq:ODE-0_R}) without loss of generality.}

\revision{One can check that for all localized structures including both DWs and vector solitons reviewed in Section \ref{sec:intro}, the background equilibria are always spatially hyperbolic in Eq.~(\ref{eq:ODE-0_R_real}). These include the $ZZ$ background of BB solitons in both the focusing Manakov system (integrable) and the focusing CNLS equation (non-integrable), the $ZN$ or $NZ$ background of DB solitons and the $NN$ background of DD solitons in both the defocusing Manakov system (integrable) and the defocusing CNLS equation (non-integrable), and so on. However, outside these well-known focusing and defocusing regimes, the spatial eigenvalues of all possible equilibria must be calculated using Eq.~(\ref{eq:char-poly}) to determine the possible existence of various localized structures.}

\section{Domain walls}\label{sec:exist-dw}

First, we seek DWs as heteroclinic orbits between two equilibria in Eq.~(\ref{eq:ODE-0_R_real}). Such an orbit requires a transverse intersection between the 2D invariant manifolds of the two equilibria in the 4D phase space, which generally exists on a codimension-1 subset of the parameter space. Thus, a heteroclinic orbit typically exists at a single value of a bifurcation parameter known as the Maxwell point \cite{knobloch2015spatial}. The Maxwell point of Eq.~(\ref{eq:ODE-0_R_real}) can be determined analytically thanks to the existence of a real-valued conserved quantity that is the Hamiltonian if Eq.~(\ref{eq:ODE-0_R_real}) is considered as a mechanical system. For later discussions, we provide the following expression of the Hamiltonian applicable to Eq.~(\ref{eq:ODE-0_R}), namely the complex version of Eq.~(\ref{eq:ODE-0_R_real}):
\begin{equation}\label{Eq:hetero_quantity}
%\begin{aligned}
    \mathcal{E}=\frac{d_1}{g_2}|\phi'|^2+\frac{\bomega_A}{g_2}|\phi|^2+\frac{g_1}{2g_2}|\phi|^4 + \frac{d_2}{g_4}|\psi'|^2+\frac{\bomega_B}{g_4}|\psi|^2+\frac{g_3}{2g_4}|\psi|^4+|\phi|^2|\psi|^2.
%\end{aligned}
\end{equation}
It can be verified that $\mathcal{E}$ is constant along any trajectory of Eq.~(\ref{eq:ODE-0_R}), namely $\mathcal{E}'=0$ assuming Eq.~(\ref{eq:ODE-0_R}). In this Hamiltonian, the two terms containing first derivatives form the kinetic energy, while the remaining terms form the potential energy. The Maxwell point between two equilibria $i$ and $j$ is then determined by
\begin{equation}\label{eq:hetero_ij}
\mathcal{E}^i=\mathcal{E}^j,
\end{equation}
where $\mathcal{E}^{i}$ represents Eq.~(\ref{Eq:hetero_quantity}) evaluated at equilibrium $i$. We note that in Ref.~\cite{coen2001domain}, the Maxwell point between two equilibria is determined by the potential energy alone, but the inclusion of the kinetic energy in Eq.~(\ref{Eq:hetero_quantity}) becomes essential if the Maxwell point needs to be determined between $\zeta$-dependent states such as plane waves.

We have checked whether Eq.~(\ref{eq:hetero_ij}) can be satisfied for each pair of equilibria in Eq.~(\ref{Eq:equilibria}) in any parameter regime, and found that the only possibilities are either between $ZZ$ and $NN$ or between $ZN$ and $NZ$.

\begin{figure}
  \centering
    \includegraphics[width=.49\textwidth]{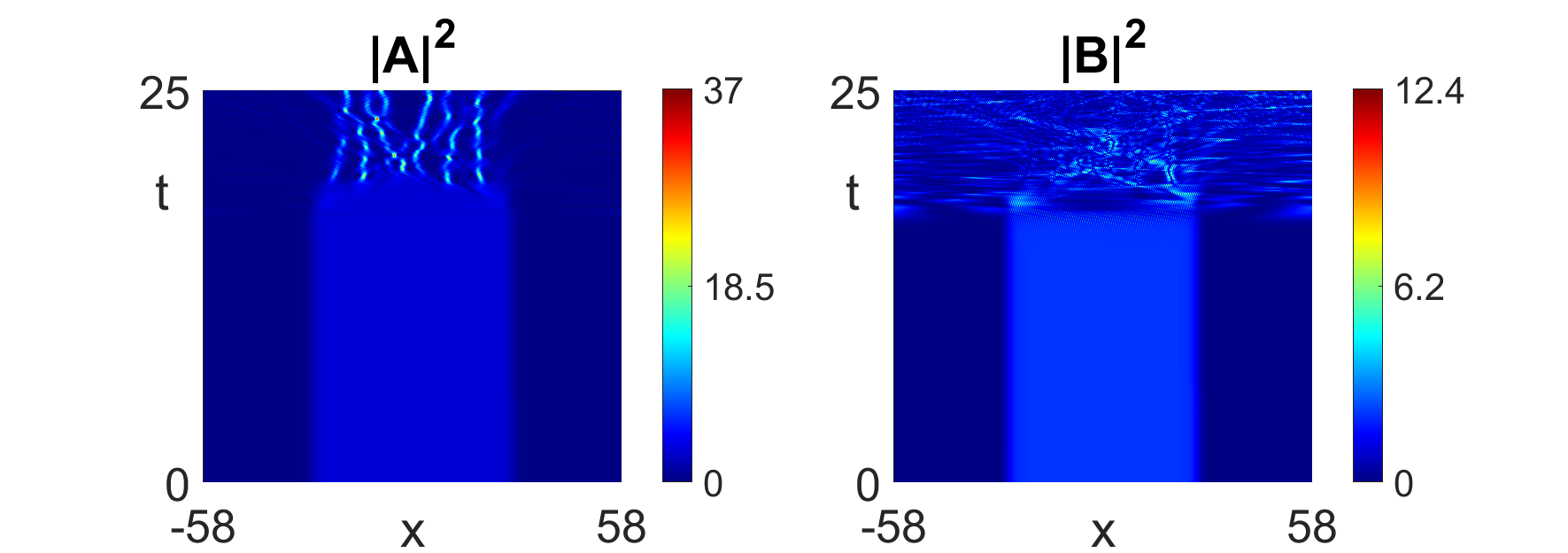}
    \caption{Time evolution in Eq.~(\ref{Eq:GCNLS}) of two asymmetric DWs between $ZZ$ and $NN$ assembled back-to-back plotted in the comoving frame. The CNLS parameters are $d_1=d_2=-5$, $g_1=-1$, $g_2=1$, $g_3=4$, and $g_4=-5$, and the travelling wave parameters are $(C_g,\omega_A,\omega_B)=(5.408,-0.5,5.5)$.}
    \label{fig:ZZ_NN_solutions}
\end{figure}

The Maxwell point between $ZZ$ and $NN$ is remarkable since $ZZ$ and $NN$ are not related by any symmetry. We shall refer to such DWs as ``asymmetric domain walls'' and discuss them qualitatively as part of our classification program. An example time evolution in Eq.~(\ref{Eq:GCNLS}) is shown in Fig.~\ref{fig:ZZ_NN_solutions} with the initial condition being two asymmetric DWs assembled back-to-back to satisfy the periodic boundary conditions in $x$. We observe that the asymmetric DW is a valid solution of Eq.~(\ref{Eq:GCNLS}) and persists for a short time before instability develops. \revision{We stress that Fig.~\ref{fig:ZZ_NN_solutions} only provides a preliminary example of asymmetric DWs, and we cannot rule out the possible existence of stable asymmetric DWs with other parameter choices. A systematic study of asymmetric DWs and related solutions is left as future work.}

By contrast, the Maxwell point between $ZN$ and $NZ$ are more natural since $ZN$ and $NZ$ can be related by exchanging $\phi$ and $\psi$ in Eq.~(\ref{eq:ODE-0_R}), although this exchange symmetry is inexact for generic parameter choices. In this case, Eq.~(\ref{eq:hetero_ij}) yields the following condition for the existence of heteroclinic orbits between $ZN$ and $NZ$:
\begin{equation}\label{eq:het_con}
\bomega=\pm\sqrt{g_3g_4/(g_1g_2)}.
\end{equation}
This heteroclinic condition defines four points on the unit circle in the $(\bomega_A,\bomega_B)$-plane, but the existence conditions for both $ZN$ and $NZ$ in Eq.~(\ref{Eq:equilibria}) choose a unique point depending on the signs of $g_1$ and $g_3$. Moreover, the spatially hyperbolic conditions for both $ZN$ and $NZ$ require the defocusing conditions $d_1g_1<0$ and $d_2g_3<0$ and the immiscibility condition $d_1d_2(g_2g_4-g_1g_3)>0$. The latter reduces to the immiscibility condition in the BEC setting when $d_1d_2>0$ \cite{trippenbach2000structure}. In this setting, DWs are well studied both physically \cite{coen2001domain} and mathematically \cite{alama2015domain}.

To complete the classification of DWs between $ZN$ and $NZ$, we express the heteroclinic condition in Eq.~(\ref{eq:het_con}) in the original traveling wave parameter space $(C_g,\omega_A,\omega_B)$ using the Galilean transformation in Eq.~(\ref{eq:galilean}). Since Eq.~(\ref{eq:het_con}) defines a half-line from the origin in the $(\bomega_A,\bomega_B)$-plane, Eqs.~(\ref{eq:galilean},\ref{eq:het_con}) define a 2D angular sector denoted by $\mathcal{A}$ in the 3D $(C_g^2,\omega_A,\omega_B)$-space where $C_g^2\geq0$ by definition. For generic choices of the CNLS parameters $(d_{1-2},g_{1-4})$, the sector $\mathcal{A}$ has a 2D projection onto the $(\omega_A,\omega_B)$-plane, so there is a unique Maxwell point with $C_g^2$ as the bifurcation parameter and $(\omega_A,\omega_B)$ fixed:
\begin{equation}\label{Eq:hetero_condition}
C_g^2=\frac{4d_1d_2\left(\omega_B\pm\omega_A\sqrt{g_3g_4/(g_1g_2)}\right)}{d_1\pm d_2\sqrt{g_3g_4/(g_1g_2)}}.
\end{equation}
However, for special choices of $(d_{1-2},g_{1-4})$ satisfying
\begin{equation}\label{eq:special_Cg}
d_1^2g_1g_2=d_2^2g_3g_4,
\end{equation}
the sector $\mathcal{A}$ is perpendicular to the $(\omega_A,\omega_B)$-plane, so $(\omega_A,\omega_B)$ must satisfy Eq.~(\ref{eq:het_con}) with $\bomega$ replaced by $\omega$, and $C_g^2\geq0$ is arbitrary at fixed $(\omega_A,\omega_B)$. This applies most notably to symmetric CNLS ($d_1=d_2$, $g_1=g_3$, $g_2=g_4$) \cite{haelterman1994polarization,haelterman1994vector,malomed1994optical,kevrekidis2004families,dror2011domain}, and more generally to a codimension-1 subset of the CNLS parameter space defined by Eq.~(\ref{eq:special_Cg}).

We stress that asymmetric DWs between $ZZ$ and $NN$ and symmetric DWs between $ZN$ and $NZ$ cannot be related by continuous deformation of parameters, namely they are homotopically nonequivalent. To see this, we first define the focusing and defocusing regimes for Eq.~(\ref{eq:ODE-0_R_real}) with general coefficients. At fixed parameters, we call Eq.~(\ref{eq:ODE-0_R_real}) focusing if the $ZZ$ equilibrium is spatially hyperbolic, and defocusing if at least one of the two equilibria, $ZN$ and $NZ$, is spatially hyperbolic. By this definition, Eq.~(\ref{eq:ODE-0_R_real}) may be focusing, defocusing, or neither, but not both. Thus, asymmetric DWs only exist in the focusing regime, while symmetric DWs only exist in the defocusing regime. Note that prior work on DWs between $ZN$ and $NZ$ in binary BECs define asymmetric and symmetric DWs depending on the symmetry of their spatial profiles \cite{filatrella2014domain,malomed2022new}. In our language, both are symmetric DWs that are typically homotopically equivalent.

After classifying heteroclinic orbits in Eq.~(\ref{eq:ODE-0_R}), the next question is whether these orbits persist into Eq.~(\ref{Eq:ODE}) with the restriction in Eq.~(\ref{eq:restri}) removed. Since equilibria of Eq.~(\ref{Eq:ODE}) correspond to plane waves in Eq.~(\ref{eq:ODE-0_R}), the Maxwell point between two equilibria in Eq.~(\ref{Eq:ODE}) can be determined as the Maxwell point between two plane waves in Eq.~(\ref{eq:ODE-0_R}) using the Hamiltonian in Eq.~(\ref{Eq:hetero_quantity}). Thus, one may naively expect the existence of a heteroclinic orbit between the $ZN$ and $NZ$ equilibria in Eq.~(\ref{Eq:ODE}) at this Maxwell point. However, the time evolution in Eq.~(\ref{Eq:GCNLS}) of an initial step between arbitrary $ZN$ and $NZ$ plane waves typically yields a DW with two dispersive shock waves respectively to the left and to the right of the DW, and this DW does satisfy the restriction in Eq.~(\ref{eq:restri}) even if the initial condition does not. A proper treatment of this time evolution requires modulation theory \cite{el2016dispersive,el2017dispersive}, but here we tentatively conclude that the restriction in Eq.~(\ref{eq:restri}) is indispensable for the existence of DWs.

\section{Vector solitons}\label{sec:vec-sol}

In this section, we first seek vector solitons as homoclinic orbits to an equilibrium in the real ODE in Eq.~(\ref{eq:ODE-0_R_real}), and then extend this solution family into the complex ODE in Eq.~(\ref{Eq:ODE}). Since Eq.~(\ref{eq:ODE-0_R_real}) is reversible under the action $R_{s_A,s_B}$ in Eq.~(\ref{eq:rev-sym-real}), the 4D phase space $(\phi,\psi,\phi',\psi')$ of Eq.~(\ref{eq:ODE-0_R_real}) is symmetric with respect to four 2D sections $\mathrm{fix}(R_{s_A,s_B})$ with $s_{A,B}=\pm$ specified by two conditions:
\begin{equation}\label{eq:boun-cond-real}
\begin{cases}
    \phi=0, & s_A=+\\
    \phi'=0, & s_A=-
\end{cases},\quad\textrm{and}\quad
\begin{cases}
    \psi=0, & s_B=+\\
    \psi'=0, & s_B=-
\end{cases}.
\end{equation}
A homoclinic orbit to an equilibrium requires a transverse intersection between the 2D stable manifold of the equilibrium and a 2D section $\mathrm{fix}(R_{s_A,s_B})$. This intersection generally exists on a codimension-0 subset of the parameter space, so any spatially hyperbolic equilibrium may support a homoclinic orbit \cite{champneys1998homoclinic}. In principle, all possible homoclinic orbits to an equilibrium can be obtained by computing the 2D stable manifold of this equilibrium, but invariant manifolds are generally hard to compute due to their complicated geometry. In practice, homoclinic orbits are often found using numerical continuation in a bifurcation parameter starting from an initial guess.

The initial guess for numerical continuation could be an analytical ansatz, but a more systematic approach is to identify bifurcation points where a solution branch originates or terminates. Thus, our analysis centers upon three codimension-1 mechanisms to generate vector solitons in Eq.~(\ref{eq:ODE-0_R_real}) from local and global bifurcation theory.

\subsection{Heteroclinic cycles}

Perhaps the most important type of global bifurcation for homoclinic orbits is the Maxwell point for heteroclinic orbits in Section \ref{sec:exist-dw}, since a heteroclinic orbit and its symmetric partner can be assembled back-to-back to form a special type of homoclinic orbit known as a heteroclinic cycle \cite{knobloch2015spatial}. A heteroclinic orbit from equilibrium $i$ to equilibrium $j$, denoted by $i\rightarrow j$, has an even partner $j\rightarrow i$ and an odd partner $-j\rightarrow-i$, for either component. A heteroclinic orbit can always be assembled with its even partner to form a heteroclinic cycle $i\rightarrow j\rightarrow i$, but it can only be assembled with its odd partner to form a heteroclinic cycle $i\rightarrow\pm j\rightarrow-i$ when $j=Z$ since $Z=-Z$, not when $j=N$ since $N\neq-N$.

Two asymmetric DWs between $ZZ$ and $NN$ can form a heteroclinic cycle with $NN$ embedded in $ZZ$, which is an extended BB soliton, or a heteroclinic cycle with $ZZ$ embedded in $NN$, which is an extended DD soliton. The extended BB soliton has multiplicity 1 and is always even-even; see the initial condition of Fig.~\ref{fig:ZZ_NN_solutions} for an example. The extended DD soliton has multiplicity 4 and can be even-even, even-odd, odd-even, or odd-odd.

Two symmetric DWs between $ZN$ and $NZ$ can form a heteroclinic cycle with $NZ$ embedded in $ZN$, or one with $ZN$ embedded in $NZ$, both of which are extended DB solitons. An extended DB soliton has multiplicity 2 and can be even-even or odd-even.

Overall, an extended vector soliton with $N_D$ dark components has multiplicity $2^{N_D}$ since a dark component can be even or odd. Besides, this multiplicity persists when the extended vector soliton is numerically continued, at least near the Maxwell point. Hence, we shall refer to a vector soliton thus obtained near the Maxwell point as a quasi-extended vector soliton. Such a vector soliton can inherit certain properties from the underlying DW even if its spatial profile is not fully extended.

\subsection{Local bifurcations}

In addition to global bifurcations that only occur in certain parameter regimes, e.g., the immiscible regime, Eq.~(\ref{eq:ODE-0_R_real}) also exhibits local bifurcations that can occur in virtually all parameter regimes. A codimension-1 local bifurcation occurs when an equilibrium becomes spatially hyperbolic, namely one or two pairs of spatial eigenvalues given by Eq.~(\ref{eq:char-poly}) leave the imaginary axis, as a bifurcation parameter, say $\bomega$, varies. A codimension-2 bifurcation occurs when two codimension-1 bifurcations collide, or the nature of a codimension-1 bifurcation changes, e.g., from supercritical to subcritical, as two bifurcation parameters, say $\bomega$ and one of $(d_{1-2},g_{1-4})$, vary.

There are two types of codimension-1 local bifurcations that can yield homoclinic orbits \cite{champneys1998homoclinic}. The first type is the Turing or Hamiltonian-Hopf bifurcation, where four purely imaginary eigenvalues collide pairwise on the imaginary axis to yield a complex quartet of eigenvalues. The second type is the pitchfork bifurcation, where two purely imaginary eigenvalues collide at the origin to yield two real eigenvalues. Note that the pitchfork bifurcation reflects the up-down symmetry of Eq.~(\ref{eq:ODE-0_R_real}), and a saddle-node bifurcation would occur without such symmetries.

\begin{figure}
  \centering
    \includegraphics[width=.49\textwidth]{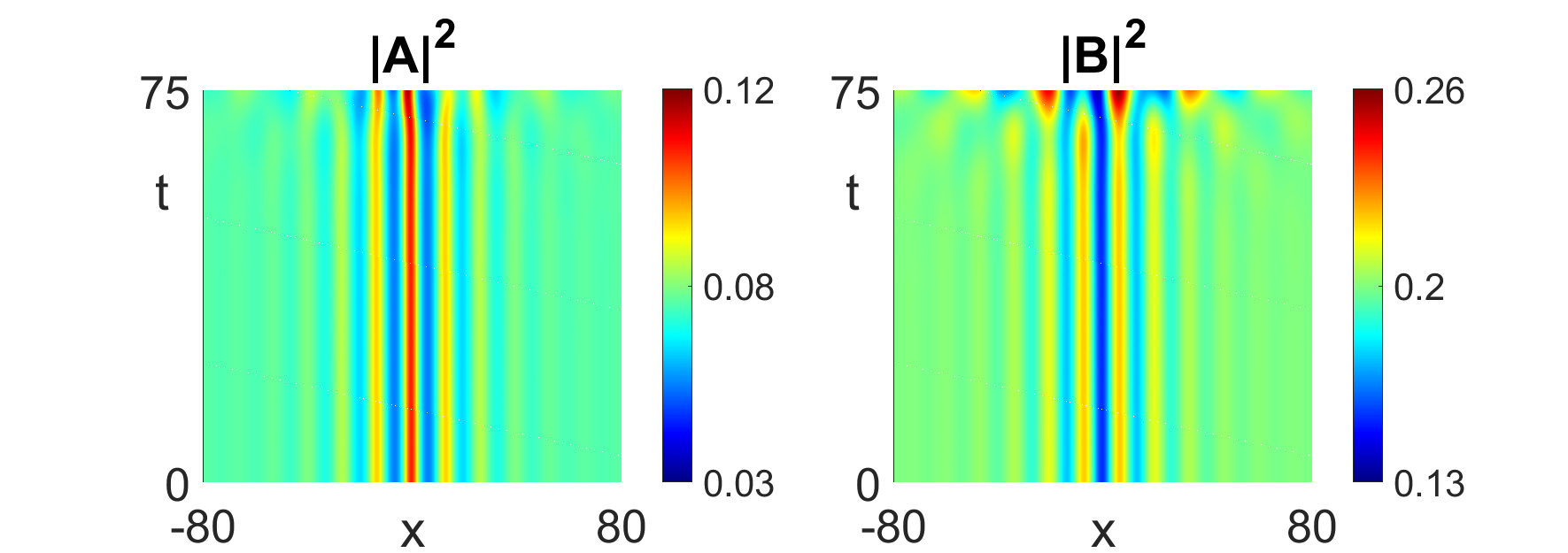}
    \caption{Time evolution in Eq.~(\ref{Eq:GCNLS}) of a Turing soliton plotted in the comoving frame. The CNLS parameters are $d_1=-2$, $d_2=-6$, $g_1=-7$, $g_2=-3$, $g_3=1$, and $g_4=6$, and the traveling wave parameters are $(C_g,\omega_A,\omega_B)=(9,-9.03,-4)$.}
    \label{fig:Turing_sols}
\end{figure}

The defining feature of the Turing bifurcation is its ability to generate spatially periodic states. As such, it has played a key role for pattern formation in nonlinear dissipative PDEs for decades. More recently, the subcritical Turing bifurcation has been much studied since it can generate spatially localized states with rich bifurcation structures \cite{knobloch2015spatial}. We have found that a subcritical Turing bifurcation can occur in Eq.~(\ref{eq:ODE-0_R_real}), but only on the $NN$ equilibrium. This bifurcation creates two branches of even-even localized states whose spatial profiles relative to $NN$ differ by an overall sign to leading nontrivial order. We shall refer to localized states near this bifurcation as weakly nonlinear ``Turing solitons'' and discuss them qualitatively as part of our classification program. An example time evolution in Eq.~(\ref{Eq:GCNLS}) is shown in Fig.~\ref{fig:Turing_sols} with the initial condition being a weakly nonlinear Turing soliton. We observe that the Turing soliton is a valid solution of Eq.~(\ref{Eq:GCNLS}) and persists for a short time before instability develops. \revision{We stress that Fig.~\ref{fig:Turing_sols} only provides a preliminary example of Turing solitons, and we cannot rule out the possible existence of stable Turing solitons with other parameter choices. A systematic study of Turing solitons and related solutions is left as future work.}

Generally, we define a Turing soliton as any vector soliton embedded in $NN$ with an oscillatory tail due to the complex spatial eigenvalues of $NN$. A component of a Turing soliton can be called dark if the center is a dip or antidark if the center is a hump. For example, the initial condition of Fig.~\ref{fig:Turing_sols} can be called a DAD Turing soliton since the center of the $A$ ($\phi$) component is a hump and the center of the $B$ ($\psi$) component is a dip. However, apart from the central hump or dip, a DAD Turing soliton has an infinite number of humps and dips in its oscillatory tail. If the DAD Turing soliton is weakly nonlinear, then the oscillatory tail can have a larger L2-norm than the central hump or dip. By contrast, a normal DAD soliton has a monotonic tail due to the real spatial eigenvalues of $NN$, so the central hump or dip generally dominates.

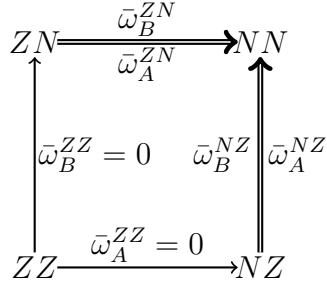
\begin{figure}
\centering
\begin{tikzpicture}
    \draw[thick,->] (0.3,0) -- (2.7,0);
    \draw[thick,->] (0,0.2) -- (0,2.8);
    \draw[thick,double,->] (3,0.2) -- (3,2.8);
    \draw[thick,double,->] (0.3,3) -- (2.7,3);
    \node at (0,0) {$ZZ$};
    \node at (3,0) {$NZ$};
    \node at (0,3) {$ZN$};
    \node at (3,3) {$NN$};
    \node at (1.5,0.3) {$\bomega_A^{ZZ}=0$};
    \node at (0.8,1.5) {$\bomega_B^{ZZ}=0$};
    \node at (1.5,2.7) {$\bomega_A^{ZN}$};
    \node at (1.5,3.3) {$\bomega_B^{ZN}$};
    \node at (2.5,1.5) {$\bomega_B^{NZ}$};
    \node at (3.5,1.5) {$\bomega_A^{NZ}$};
\end{tikzpicture}
\caption{Network of pitchfork bifurcations in Eq.~(\ref{eq:ODE-0_R_real}) with the nodes representing the four equilibria and the edges representing the bifurcation points. The two Maxwell points respectively for asymmetric and symmetric DWs would occupy the two diagonal edges, but they are not shown explicitly.} \label{fig:pitch-net}
\end{figure}

In a dynamical system with an up-down symmetry, there is always an equilibrium at zero, and the pitchfork bifurcation causes two nonzero equilibria related by the up-down symmetry to emerge from the zero equilibrium as a parameter varies. A pitchfork bifurcation in Eq.~(\ref{eq:ODE-0_R_real}) can occur on a zero component regardless of whether the other component is zero or nonzero. Thus, $ZN$ or $NZ$ can bifurcate from $ZZ$, and $NN$ can bifurcate from $ZN$ or $NZ$. For the latter, the bifurcation parameter can be chosen as either $\bomega_A$ or $\bomega_B$. However, for the former, the bifurcation parameter can only be chosen as the $\bomega_{A,B}$ that the $ZN$ or $NZ$ equilibrium depends on. Thus, there are no ambiguities if the $\bomega_{A,B}$ value where an equilibrium $i$ bifurcates from an equilibrium $j$ is denoted by $\bomega_{A,B}^j$. For example, the $\bomega_A$ value where $NN$ bifurcates from $ZN$ is denoted by $\bomega_A^{ZN}$. A network can be constructed to visualize these pitchfork bifurcations with the nodes representing the four equilibria and the edges representing the bifurcation points; see Fig.~\ref{fig:pitch-net}. Note that the two types of Maxwell points in Section \ref{sec:exist-dw} represented on this network would occupy the two diagonal edges.

Analytical expressions can be derived for weakly nonlinear vector solitons near these pitchfork bifurcation points using asymptotic analysis. The details of these rather standard calculations will be omitted, but we note that asymptotic analyses near different pitchfork bifurcations always yield the same second-order ODE, known as the normal form, only with different coefficients \cite{champneys1998homoclinic}. This normal form ODE can admit, in different parameter regimes, either an even homoclinic orbit to the zero equilibrium, or an odd homoclinic orbit connecting the two nonzero equilibria related by the up-down symmetry.

Near $\bomega_A^{ZZ}=0$ or $\bomega_B^{ZZ}=0$, one of the two components remains zero past the bifurcation point, so we obtain the pitchfork bifurcation in the scalar NLS equation that yields the well-known branches of exact solutions. The even homoclinic orbit corresponds to scalar bright solitons in the focusing regime with a sech shape, while the odd homoclinic orbit corresponds to scalar dark solitons in the defocusing regime with a tanh shape. These branches of exact solutions are not vector solitons and so will not be shown explicitly. However, one can identify codimension-1 bifurcation points on the branch of scalar dark solitons from the perspective of quantum mechanics and obtain branches of DB solitons whose bright components can have one or more humps \cite{achilleos2014beating,charalampidis2015dark}. In Section \ref{sec:exa-sol}, we shall present another application of exact solutions to scalar NLS by ``lifting'' them to codimension-1 families of exact solutions to Eq.~(\ref{eq:ODE-0_R_real}) that can be numerically continued directly into vector solitons.

The pitchfork bifurcation point on $ZN$ is given by
\begin{equation}\label{eq:pitch-pt-zn}
    \bomega^{ZN}=g_3/g_2.
\end{equation}
Near $\bomega_A^{ZN}$, we assume that $\bomega_A-\bomega_A^{ZN}=O(\epsilon^2)$, where $0<\epsilon\ll1$, and use the expansion
\begin{equation}\label{eq:pitch-ansat}
    \begin{bmatrix}
   \phi \\ \psi 
    \end{bmatrix}=
        \begin{bmatrix}
   \phi_0 \\ \psi_0 
    \end{bmatrix}+
        \begin{bmatrix}
   \tilde{\phi} \\ \tilde{\psi} 
    \end{bmatrix},
\end{equation}
where $[\phi_0,\psi_0]^T$ is the perturbed $ZN$ equilibrium, and $[\tilde{\phi},\tilde{\psi}]^T$ is the $\zeta$-dependent part that determines the spatial profile of the vector soliton. The pitchfork bifurcation is subcritical when $d_1g_3(g_2g_4-g_1g_3)<0$. In this case, the leading nontrivial order for $\tilde{\phi}$ is $O(\epsilon)$ and contains the even homoclinic orbit of the normal form, which is a bright component of the vector soliton:
\begin{equation}
    \tilde{\phi}=\sqrt{\frac{2g_3(\bomega_A-\bomega_A^{ZN})}{g_2g_4-g_1g_3}}\mbox{sech}\left(\sqrt{\frac{\bomega_A-\bomega_A^{ZN}}{-d_1}}\zeta\right).
\end{equation}
The pitchfork bifurcation is supercritical when $d_1g_3(g_2g_4-g_1g_3)>0$. In this case, the leading nontrivial order for $\tilde{\phi}$ is $O(\epsilon)$ and contains the odd homoclinic orbit of the normal form, which is a dark component of the vector soliton:
\begin{equation}\label{eq:pitch-black}
    \tilde{\phi}=\sqrt{\frac{g_3(\bomega_A-\bomega_A^{ZN})}{g_2g_4-g_1g_3}}\mbox{tanh}\left(\sqrt{\frac{\bomega_A-\bomega_A^{ZN}}{2d_1}}\zeta\right).
\end{equation}
In both the subcritical and the supercritical cases, the leading nontrivial order for $\tilde{\psi}$ is $O(\epsilon^2)$ and contains an even localized profile ``induced'' by $\tilde{\phi}$:
\begin{equation}
    \tilde{\psi}=-\frac{g_4}{2g_3\psi_0}\tilde{\phi}^2,
\end{equation}
where $\psi_0=\sqrt{-\bomega_B/g_3}$. In the subcritical case, $\tilde{\psi}$ is a dark component when $g_3g_4>0$, and an antidark component when $g_3g_4<0$. Thus, the former is a DB soliton, while the latter is a BAD soliton. In the supercritical case, $\tilde{\psi}$ is an antidark component when $g_3g_4>0$, and a dark component when $g_3g_4<0$. Thus, the former is a DAD soliton, while the latter is a DD soliton. In the BEC setting, the existence of these four types of vector solitons is established from the perspective of quantum mechanics in Ref.~\cite{kevrekidis2004families}, where the antidark concept is first introduced. In the physics community, DAD solitons are better known as magnetic solitons, which are actively studied theoretically \cite{qu2016magnetic,chai2022magnetic} and experimentally \cite{farolfi2020observation}.

The pitchfork bifurcation point on $NZ$ is given by
\begin{equation}\label{eq:pitch-pt-nz}
    \bomega^{NZ}=g_4/g_1.
\end{equation}
Near $\bomega_A^{NZ}$, we can again assume $\bomega_A-\bomega_A^{NZ}=O(\epsilon^2)$, use the expansion in Eq.~(\ref{eq:pitch-ansat}), and obtain the four types of vector solitons similarly to the $\bomega_A^{ZN}$ case. These solutions are summarized next but not shown explicitly. The pitchfork bifurcation is subcritical when $d_2g_1(g_2g_4-g_1g_3)<0$. In this case, $\tilde{\psi}$ is an $O(\epsilon)$ bright component that is even. The pitchfork bifurcation is supercritical when $d_2g_1(g_2g_4-g_1g_3)>0$. In this case, $\tilde{\psi}$ is an $O(\epsilon)$ dark component that is odd. In both cases, $\tilde{\phi}$ is an $O(\epsilon^2)$ dark or antidark component that is even. In the subcritical case, $\tilde{\phi}$ is dark if $g_1g_2>0$ and antidark if $g_1g_2<0$. These are respectively a DB soliton and a BAD soliton. In the supercritical case, $\tilde{\phi}$ is antidark if $g_1g_2>0$ and dark if $g_1g_2<0$. These are respectively a DAD soliton and a DD soliton. Notably, these DD or DAD solitons bifurcating from $\bomega_A^{NZ}$ may coexist with the DD or DAD solitons bifurcating from $\bomega_A^{ZN}$, but these two branches are distinct since they have opposite parities in both components.

The pitchfork bifurcation is the fundamental bifurcation connecting the four equilibria as visualized in Fig.~\ref{fig:pitch-net}. Thus, after analyzing the codimension-1 bifurcations, it is also useful to identify the codimension-2 bifurcations. To identify the codimension-2 pitchfork bifurcation point for any equilibrium from Fig.~\ref{fig:pitch-net}, one can first identify the node representing this equilibrium and then equate the two codimension-1 points represented by the two edges meeting at this node. Since a codimension-2 point selects a triangle in Fig.~\ref{fig:pitch-net} that contains a diagonal edge, it can provide valuable insights into the interplay between DWs and vector solitons. We shall identify the codimension-2 points but leave a detailed study as future work.

For $ZZ$, there are no nontrivial codimension-2 points.

For $NN$, Eqs.~(\ref{eq:pitch-pt-zn},\ref{eq:pitch-pt-nz}) imply that the codimension-2 point is $g_2g_4-g_1g_3=0$. This is the miscible-immiscible transition point, which separates the miscible regime featuring DD solitons and the immiscible regime featuring symmetric DWs between $ZN$ and $NZ$ \cite{ohberg2001dark}.

For $ZN$, Eq.~(\ref{eq:pitch-pt-zn}) and $\bomega_B^{ZZ}=0$ imply that the codimension-2 point is $g_3=0$. For $NZ$, Eq.~(\ref{eq:pitch-pt-nz}) and $\bomega_A^{ZZ}=0$ imply that the codimension-2 point is $g_1=0$. These are focusing-defocusing transition points, which separate the focusing regime featuring asymmetric DWs between $ZZ$ and $NN$ and the defocusing regime featuring DB solitons embedded in either $ZN$ or $NZ$.

% However, the normal form near a codimension-2 point likely takes the same form as Eq.~(\ref{eq:ODE-0_R_real}), possibly with the same coefficients, so there is no dimension reduction that simplifies the problem. Nonetheless, analysis of codimension-2 points could reveal self-similarity in the solution space of Eq.~(\ref{eq:ODE-0_R_real}), which is an interesting topic in its own right.

\subsection{Exact solutions}\label{sec:exa-sol}

As an application of exact solutions to scalar NLS including scalar bright and dark solitons, we lift them to exact solutions to Eq.~(\ref{eq:ODE-0_R_real}) by requiring a scalar soliton multiplied by a constant amplitude vector to satisfy Eq.~(\ref{eq:ODE-0_R_real}).

To find scalar bright solitons with vector amplitudes, we substitute the following ansatz into Eq.~(\ref{eq:ODE-0_R_real}):
\begin{equation}
    \begin{bmatrix}
   \phi \\ \psi 
    \end{bmatrix}=
    \begin{bmatrix}
   A_1 \\ A_2 
    \end{bmatrix}
    \mbox{sech}\left(W \zeta\right),
\end{equation}
where $A_1, A_2, W>0$ without loss of generality. This yields the following consistency condition:
\begin{equation}\label{eq:consist-exact}
    \bomega_B=\frac{d_2}{d_1}\bomega_A,
\end{equation}
while the soliton parameters are
\begin{gather*}
    W^2=-\frac{\bomega_A}{d_1},\quad A_1^2=\frac{2W^2(d_1g_3-d_2g_2)}{g_1g_3-g_2g_4},\quad
    A_2^2=\frac{2W^2(d_2g_1-d_1g_4)}{g_1g_3-g_2g_4}.
\end{gather*}

To find scalar dark solitons with vector amplitudes, we substitute the following ansatz into Eq.~(\ref{eq:ODE-0_R_real}):
\begin{equation}\label{eq:scalar-dd-ansatz}
    \begin{bmatrix}
   \phi \\ \psi 
    \end{bmatrix}=
    \begin{bmatrix}
   A_1 \\ A_2 
    \end{bmatrix}
    \mbox{tanh}\left(W \zeta\right),
\end{equation}
where $A_1, A_2, W>0$ without loss of generality. This yields the same consistency condition as the bright case given by Eq.~(\ref{eq:consist-exact}), while the soliton parameters are
\begin{gather*}
    W^2=\frac{\bomega_A}{2d_1},\quad A_1^2=\frac{2W^2(d_2g_2-d_1g_3)}{g_1g_3-g_2g_4},\quad
    A_2^2=\frac{2W^2(d_1g_4-d_2g_1)}{g_1g_3-g_2g_4}.
\end{gather*}

Arguably, scalar bright or dark solitons with vector amplitudes are not true BB or DD vector solitons. However, they are reversible homoclinic orbits that happen to have exact analytical expressions when the consistency condition in Eq.~(\ref{eq:consist-exact}) is satisfied. Once they are numerically continued such that Eq.~(\ref{eq:consist-exact}) is no longer satisfied, they become true BB or DD vector solitons whose profiles cannot be exactly a sech or a tanh function. Thus, we shall consider the exact solutions themselves as true vector solitons and call them scalar BB and DD solitons.

\subsection{Summary of real vector solitons}

After identifying the three codimension-1 mechanisms for generating vector solitons in Eq.~(\ref{eq:ODE-0_R_real}), we can group these vector solitons by their types and possible parities; see Table \ref{table:vs-type-parity}. Overall, a bright or an antidark component is always even, while a dark component can be either even or odd. Any codimension-1 family of vector solitons in Table \ref{table:vs-type-parity} can be numerically continued into a codimension-0 family of vector solitons, namely a 2-parameter family in $(\bomega_A,\bomega_B)$ or equivalently a 3-parameter family in $(C_g,\omega_A,\omega_B)$ using the Galilean transformation in Eq.~(\ref{eq:galilean}). These codimension-0 families of vector solitons may not be distinct since two vector solitons of the same type and parity but generated by different mechanisms could be parametrically related in Eq.~(\ref{eq:ODE-0_R_real}).

\begin{table}
\centering
    \begin{tabular}{| c | c |}
    \hline
    Type of vector soliton & Possible parities \\ \hhline{|==|}
    \multicolumn{2}{|c|}{{\it Heteroclinic cycles}} \\ \hline
    (Quasi-)Extended BB & 1 EE \\ \hline
    (Quasi-)Extended DD & 1 EE + 1 EO + 1 OE + 1 OO \\ \hline
    (Quasi-)Extended DB & 1 EE + 1 OE \\ \hhline{|==|}
    \multicolumn{2}{|c|}{{\it Local bifurcations}} \\ \hline
    Weakly nonlinear DB & 1 EE \\ \hline
    Weakly nonlinear BAD & 1 EE \\ \hline
    Weakly nonlinear DD & 1 EO/OE \\ \hline
    Weakly nonlinear DAD & 1 OE \\ \hline
    Weakly nonlinear Turing & 2 EE \\ \hhline{|==|}
    \multicolumn{2}{|c|}{{\it Exact solutions}} \\ \hline
    Scalar BB & 1 EE \\ \hline
    Scalar DD & 1 OO \\ \hline
    \end{tabular}
    \caption{Types of vector solitons and their possible parities. Abbreviations: EE = even-even; EO = even-odd; OE = odd-even; OO = odd-odd.}
    \label{table:vs-type-parity}
\end{table}

BAD, DAD, and Turing solitons can only be generated by local bifurcations, not heteroclinic cycles or exact solutions. Nonetheless, since Turing, DD, and DAD solitons share the $NN$ background, a Turing soliton could be parametrically related to a DD or a DAD soliton via the so-called Belyakov-Devaney point, where the spatial eigenvalues given by Eq.~(\ref{eq:char-poly}) transition from a complex quartet of eigenvalues to four real eigenvalues by colliding pairwise on the real axis, or vice versa \cite{champneys1998homoclinic}.

There are two types of BB solitons including extended and scalar BB solitons. Both are even-even, so they could be parametrically related.

There are two types of DB solitons including extended and weakly nonlinear DB solitons. The former can be even-even or odd-even, while the latter are even-even. Thus, even-even extended DB solitons could be parametrically related to weakly nonlinear DB solitons.

There are three types of DD solitons including extended, weakly nonlinear, and scalar DD solitons. Extended DD solitons can have all four possible parities. Weakly nonlinear DD solitons are either even-odd or odd-even, say even-odd. Scalar DD solitons are odd-odd. Thus, even-odd extended DD solitons could be parametrically related to weakly nonlinear DD solitons, while odd-odd extended DD solitons could be parametrically related to scalar DD solitons.

\subsection{Complex vector solitons}\label{sec:complex-vs}

Once we have a 3-parameter family of real homoclinic orbits in $(C_g,\omega_A,\omega_B)$ in Eq.~(\ref{eq:ODE-0_R_real}), we can remove the restriction in Eq.~(\ref{eq:restri}) and numerically continue these orbits in $(k_A,k_B)$ to obtain a 5-parameter family of complex homoclinic orbits in Eq.~(\ref{Eq:ODE}). Although these orbits are different solutions to the ODE in Eq.~(\ref{Eq:ODE}), they do not always yield different vector solitons in the PDE in Eq.~(\ref{Eq:GCNLS}).

Consider a numerical continuation from $k_A$ to $\hat{k}_A$; the $k_B$ case is similar. To identify which ODE solutions yield the same vector soliton in the PDE, we observe that the $A$ component of a vector soliton $A=\phi(\zeta)\exp(i(k_Ax-\omega_At))$ can be rewritten as $A=\hat{\phi}(\zeta)\exp(i(\hat{k}_Ax-\hat{\omega}_At))$ where
\begin{equation}\label{eq:phi-equiv}
    \hat{\phi}(\zeta)=\phi(\zeta)\exp(i(k_A-\hat{k}_A)\zeta)
\end{equation}
and, recalling that $\zeta=x-C_gt$,
\begin{equation}\label{eq:omega-equiv}
\hat{\omega}_A=\omega_A+C_g(\hat{k}_A-k_A).
\end{equation}
Thus, the two solution families, $\hat{\phi}$ at $\hat{k}_A$ and $\phi$ at $k_A$, yield the same family of vector solitons.

If $\phi$ is homoclinic to the $N$ equilibrium, namely dark or antidark, then $\hat{\phi}$ is homoclinic to a nonzero plane wave. Since the numerical continuation from $k_A$ to $\hat{k}_A$ yields a solution family homoclinic to the $N$ equilibrium, this solution family cannot be equivalent to $\hat{\phi}$, so this continuation must yield a different family of vector solitons. Notably, this continuation might connect two branches of real vector solitons with different parities.

However, if $\phi$ is homoclinic to the $Z$ equilibrium, namely bright, then $\hat{\phi}$ is also homoclinic to the $Z$ equilibrium. Since the numerical continuation from $k_A$ to $\hat{k}_A$ also yields a solution family homoclinic to the $Z$ equilibrium, this solution family must be equivalent to $\hat{\phi}$ with the equivalence given by Eqs.~(\ref{eq:phi-equiv}--\ref{eq:omega-equiv}), so this continuation does not yield a different family of vector solitons.

\revision{The nontrivial continuation in the wavenumber for a dark or antidark component, but not a bright component, is perhaps best known in the scalar NLS equation, where scalar bright solitons have two parameters including the group velocity and the frequency, while scalar dark solitons have three parameters including the group velocity, the frequency, and the wavenumber \cite{Ablowitz-2013}. The profile of a scalar dark soliton is a real tanh function plus an imaginary constant $i\nu$, where $\nu$ is the darkness parameter with $\nu=0$ for a black soliton and $\nu\neq0$ for a gray soliton. In our language, a black soliton is a real scalar dark soliton, a gray soliton is a complex scalar dark soliton, and the former can be continued in the wavenumber to obtain the latter.}

\revision{The continuation in the wavenumber of a dark or antidark component from real to complex can always be performed numerically, but a complex dark component may also be expressible analytically as exemplified by the scalar NLS equation. In the CNLS equation, one can similarly generalize a black component described by a real tanh function to a gray component by adding an imaginary constant. There are three types of real vector solitons in Table \ref{table:vs-type-parity} with such a black component including weakly nonlinear DD, weakly nonlinear DAD, and scalar DD solitons. For the first two types, the $O(\epsilon)$ component in Eq.~(\ref{eq:pitch-black}) can be generalized from black to gray similarly to scalar NLS, which will not be shown explicitly. For the third type, both components can be generalized from black to gray by adding an imaginary vector to Eq.~(\ref{eq:scalar-dd-ansatz}):
\begin{equation}\label{eq:complex-dd-ansatz}
    \begin{bmatrix}
   \phi \\ \psi 
    \end{bmatrix}=
    \begin{bmatrix}
   A_1 \\ A_2 
    \end{bmatrix}
    \mbox{tanh}\left(W \zeta\right)+i
    \begin{bmatrix}
   B_1 \\ B_2 
    \end{bmatrix},
\end{equation}
where $A_1, A_2, W>0$ without loss of generality, and $B_1,B_2\in\mathbb{R}$. Substituting this ansatz into Eq.~(\ref{Eq:ODE}) with the following shorthand notations for its coefficients
\begin{equation*}
    K_A\equiv2d_1k_A-C_g, \quad \Omega_A\equiv\omega_A-d_1k_A^2, \quad K_B\equiv2d_2k_B-C_g, \quad \Omega_B\equiv\omega_B-d_2k_B^2,
\end{equation*}
we obtain the following consistency condition
\[
\Omega_B = \frac{(d_2 g_2 - d_1 g_3) (d_2 g_1 - d_1 g_4) (d_2^2 K_A^2 - d_1^2 K_B^2)}{2 d_1^3 d_2^2 (g_1 g_3 - g_2 g_4)} + \frac{d_2 \Omega_A}{d_1},
\]
while the soliton parameters are
\[
   W^2 = \frac{d_2 g_1 g_2 (d_2^2 K_A^2 - d_1^2 K_B^2) - d_1 d_2^2 g_1 g_3 K_A^2 + d_1^3 g_2 g_4 K_B^2}{4 d_1^3 d_2^2 (g_1 g_3 - g_2 g_4)}+\frac{\Omega_A}{2 d_1},
\]
   \[
   A_1^2 = \frac{2 (d_2 g_2 - d_1 g_3) W^2}{g_1 g_3 - g_2 g_4},\quad   A_2^2 = \frac{2 (d_1 g_4 - d_2 g_1) W^2}{g_1 g_3 - g_2 g_4},
   \]
   \[
   B_1 = \frac{K_AA_1W}{g_1A_1^2+g_2A_2^2},\quad   B_2 = \frac{K_BA_2W}{g_3A_2^2+g_4A_1^2}.
   \]
Thus, the 2-parameter family of black-black solitons in Eq.~(\ref{eq:scalar-dd-ansatz}) can be generalized to the 4-parameter family of gray-gray solitons in Eq.~(\ref{eq:complex-dd-ansatz}). The two darkness parameters $B_1$ and $B_2$ become nonzero respectively when $K_A$ and $K_B$ become nonzero, so this generalization is essentially a continuation in $(K_A,K_B)$, or equivalently the two wavenumbers $(k_A,k_B)$. The general family of complex DD solitons have five parameters and can be obtained from this family of gray-gray solitons by numerical continuation in $\Omega_B$, or equivalently $\omega_B$, such that the above consistency condition is violated.
}

Although a bright component cannot be non-trivially continued in the wavenumber like a dark or an antidark component, the latter cannot be non-trivially multiplied by a constant phase factor $e^{i\Theta_A}$ or $e^{i\Theta_B}$ like the former for vector soliton collisions. This is because in a collision between two vector soliton respectively located on the left and on the right, the left and right dark or antidark components must have matching phases, while the left and right bright components can have any phases. Moreover, the left and right bright components can also have different frequencies, so one can expect richer dynamics when a soliton collision involves at least one bright component. If the $A$ component of a vector soliton is bright, then we define the polarization of this vector soliton as the phase difference $\rho=\Theta_A-\Theta_B$ regardless of whether the $B$ component is bright, dark, or antidark.

\section{Numerical continuation}\label{sec:num-con}

Since vector solitons are reversible homoclinic orbits, only half of the soliton profile needs to be numerically continued, say the right half. Thus, we can choose a domain $\zeta\in[0,L]$ with $L$ large, such that $\zeta=0$ is the center of the soliton and $\zeta\rightarrow L$ is the tail of the soliton. At $\zeta=L$, we can impose Neumann boundary conditions such that the tail decays to the background equilibrium. At $\zeta=0$, we must impose boundary conditions required by the reversibility symmetry of the soliton in question.

To numerically continue a real vector soliton with the reversibility in Eq.~(\ref{eq:rev-sym-real}), we must impose at $\zeta=0$ the boundary conditions in Eq.~(\ref{eq:boun-cond-real}), which depend on the parities of the two components. As stated earlier, to convert a real vector soliton to a complex vector soliton with the reversibility $R_{0,0}$ in Eq.~(\ref{eq:rev-sym}), any odd component must be multiplied by $i$. To numerically continue a complex vector soliton with the reversibility $R_{0,0}$, we must impose at $\zeta=0$ the boundary conditions corresponding to the symmetric section $\mathrm{fix}(R_{0,0})$ given explicitly by
\begin{equation}\label{eq:rev-bou-con}
    \mathcal{I}\mbox{m}(\phi,\psi)=0,\quad\mathcal{R}\mbox{e}(\phi',\psi')=0.
\end{equation}

As an application of our classification program, we consider an example with the CNLS parameters $d_1=-1$, $d_2=-3$, $g_1=2$, $g_2=4$, $g_3=5$, and $g_4=8$. This parameter combination is in the defocusing regime relevant to BECs, and the immiscibility condition $g_2g_4-g_1g_3>0$ is satisfied. Thus, the $ZN$ and $NZ$ equilibria can be both spatially hyperbolic, and a DW may form at the Maxwell point in Eq.~(\ref{Eq:hetero_condition}). An example combination of the traveling wave parameters satisfying both conditions is $(C_g,\omega_A,\omega_B)=(5.093,-9.9,-9.8)$. To compute a DW profile at the Maxwell point, we make Eq.~(\ref{eq:ODE-0_R_real}) the steady state of a dissipative PDE and evolve an initial step between $ZN$ and $NZ$ in this PDE to reach the steady state, but there are other numerical methods available.

\begin{figure}
  \centering
      \subfigure[]{\includegraphics[width=.49\textwidth]{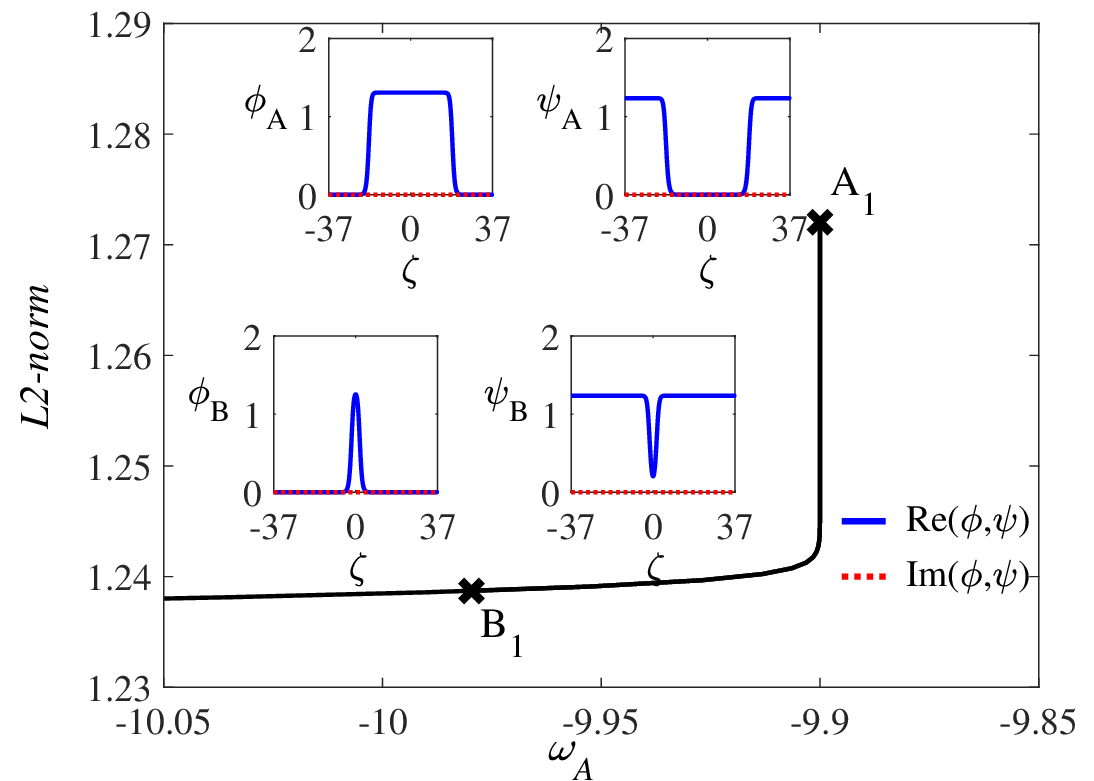}\label{fig:omegaA_bif}}
      \subfigure[]{\includegraphics[width=.49\textwidth]{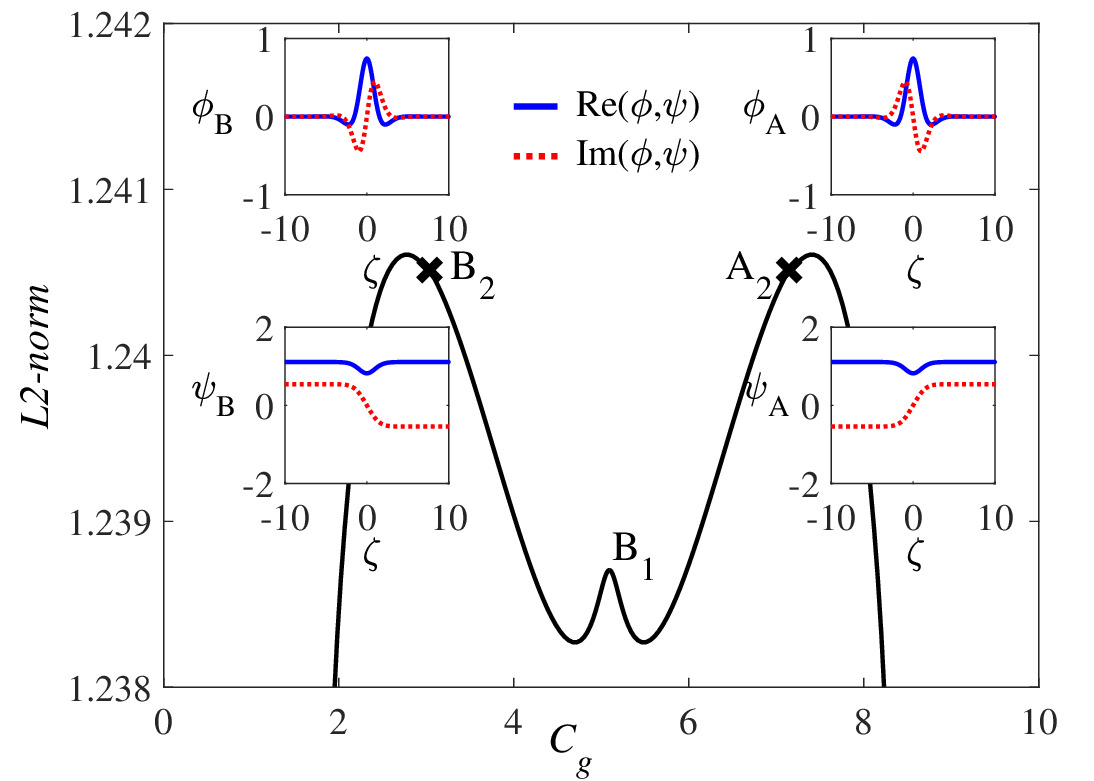}\label{fig:Cg_bif}}
    \caption{(Color online) Bifurcation diagrams and the profiles of the accompanying solutions for numerical continuations in $\omega_A$ shown in panel (a), followed by $C_g$ shown in panel (b). The solution profiles are shown in terms of their real (blue-solid) and imaginary (red-dashed) parts. The CNLS parameters are $d_1=-1$, $d_2=-3$, $g_1=2$, $g_2=4$, $g_3=5$, and $g_4=8$, and the traveling wave parameters at the starting point $A_1$ in panel (a) is $(C_g,\omega_A,\omega_B)=(5.093,-9.9,-9.8)$. (a) The profile at $A_1$ is shown as $\phi_A$ and $\psi_A$, and the profile at $B_1$ is shown as $\phi_B$ and $\psi_B$. (b) The profile at $A_2$ is shown as $\phi_A$ and $\psi_A$, and the profile at $B_2$ is shown as $\phi_B$ and $\psi_B$.}
    \label{fig:bif_diagrams}
\end{figure}

Once a DW profile between $ZN$ and $NZ$ is computed, it can be treated as half of the profile of an extended DB soliton and numerically continued into a codimension-0 family of real vector solitons. As listed in Table \ref{table:vs-type-parity}, an extended DB soliton has two possible parities including even-even and odd-even. We shall focus on the former and leave the latter as future work. Thus, we impose even-even boundary conditions at $\zeta=0$, namely Eq.~(\ref{eq:boun-cond-real}) with $s_A=s_B=+$, Neumann boundary conditions at $\zeta=L$, and numerically continue the DW in one of the traveling wave parameters $(C_g,\omega_A,\omega_B)$, say $\omega_A$. We use the software AUTO-07p \cite{doedel2007auto} for all numerical continuations in this paper, but other software can also be used. The bifurcation diagram is shown in Fig.~\ref{fig:omegaA_bif} with the profiles of the accompanying solutions shown in the insets. Note that although the full soliton profile is plotted, only half of the profile is numerically continued. The starting point is the extended DB soliton at $A_1$ whose profile is shown as $\phi_A$ and $\psi_A$ and consists of two DWs. As $\omega_A$ varies, the two DWs ``merge'' into a generic real DB soliton at $B_1$ whose profile is shown as $\phi_B$ and $\psi_B$. Although not shown explicitly in Fig.~\ref{fig:omegaA_bif}, this branch of even-even DB solitons terminates at the pitchfork bifurcation point $\omega_A^{ZN}$ on the $ZN$ equilibrium.

As discussed in Section \ref{sec:complex-vs}, a 3-parameter family of real DB solitons in $(C_g,\omega_A,\omega_B)$ can be numerically continued in the wavenumber of the dark component $k_B$ into a 4-parameter family of complex DB solitons. A key motivation to compute complex vector solitons is to study interactions (collisions) involving such solitons, during which the interacting solitons must share the same plane wave background. Thus, we impose reversible boundary conditions in Eq.~(\ref{eq:rev-bou-con}) at $\zeta=0$, Neumann boundary conditions at $\zeta=L$, and numerically continue complex DB solitons in $C_g$ with $k_B$ fixed rather than in $k_B$ with $C_g$ fixed. The bifurcation diagram is shown in Fig.~\ref{fig:Cg_bif} with the profiles of the accompanying solutions shown in the insets. The starting point is the real DB soliton at $B_1$ in Fig.~\ref{fig:omegaA_bif}. As $C_g$ increases, one obtains a complex DB soliton at $A_2$ whose profile is shown as $\phi_A$ and $\psi_A$. As $C_g$ decreases, one obtains a complex DB soliton at $B_2$ whose profile is shown as $\phi_B$ and $\psi_B$. The bifurcation diagram is symmetric with respect to $B_1$ and the solution profiles at $A_2$ and $B_2$ are related by complex conjugation. This is expected since Eq.~(\ref{Eq:ODE}) to the left and to the right of $B_1$ differ only in the sign of the coefficient of $\psi'$.

Extended DB solitons formed by two DWs are well studied in the optics setting where they are often called domain wall solitons \cite{haelterman1994polarization,haelterman1994vector,malomed1994optical,jovanoski2008exact}, and in the BEC setting where they are sometimes called bubble droplets \cite{filatrella2014domain}. The above example shows that DWs and DB solitons can be parametrically related in the immiscible regime of the asymmetric CNLS equation relevant to BECs. An interesting prospect is to deduce analytically the properties of quasi-extended DB solitons near the Maxwell point from the properties of the underlying DW.

\section{Collisions between two polarized dark-bright solitons}\label{sec:pdb-collision}

In Section \ref{sec:num-con}, we have obtained a 4-parameter family of complex DB solitons in $(C_g,\omega_A,\omega_B,k_B)$ with the $B$ component being dark and the $A$ component being bright. As discussed in Section \ref{sec:complex-vs}, the bright component $A$ can be multiplied by a constant phase factor $e^{i\Theta_A}$ to yield a 5-parameter family of polarized DB solitons in $(C_g,\omega_A,\omega_B,\Theta_A,k_B)$ for soliton collisions. In such collisions, the 2 parameters $(\omega_B,k_B)$ must be the same between the two colliding solitons such that the solitons share the same plane wave background, but the other 3 parameters $(C_g,\omega_A,\Theta_A)$ can be different. Among these, $C_g$ must be different for the collision to occur, $\omega_A$ will be taken as the same for simplicity, and we shall focus on the effect of different $\Theta_A$. We shall denote $\Theta_A$ as the polarization $\rho$ recalling that $\rho=\Theta_A-\Theta_B$ and $\Theta_B=0$ for the dark component $B$. In the BEC setting, it is well known that collisions between two polarized DB solitons can yield a mass exchange between the two bright components \cite{busch2001dark}. We shall examine this mass exchange in the immiscible regime through a numerical example.

To reduce radiation for soliton collisions, we consider an example with the CNLS parameters $d_1=d_2=-1$, $g_1=g_2=g_3=2$, and $g_4=3$, which differs from the defocusing Manakov system only by adding 1 to $g_4$. Apart from near integrability, this example and the example in Section \ref{sec:num-con} have similar bifurcation structures since both belong to the immiscible regime. A DW exists at the Maxwell point $(C_g,\omega_A,\omega_B)=(5.395,-9.5,-10)$. A real DB soliton can be found by numerical continuation of the DW to $\omega_A=-9.7$ similarly to Fig.~\ref{fig:omegaA_bif}. A family of complex DB solitons sharing the same plane wave background can be obtained by numerical continuation of the real DB soliton in $C_g$ similarly to Fig.~\ref{fig:Cg_bif}. These two bifurcation diagrams will not be shown explicitly since they are similar to Fig.~\ref{fig:bif_diagrams}(a,b). We shall call the group velocity of the real DB soliton $C_g^{(0)}=5.395$. For soliton collisions, we shall choose two complex DB solitons with $C_g$ values symmetric with respect to $C_g^{(0)}$ similarly to Fig.~\ref{fig:Cg_bif}, which are respectively called $(C_g^{(1)},C_g^{(2)})=(5.789,5)$. Thus, these two solitons have opposite velocities in the frame comoving with velocity $C_g^{(0)}$, and the two soliton profiles are related by complex conjugation.

\begin{figure}
  \centering
  \subfigure[]{\includegraphics[width=.49\textwidth]{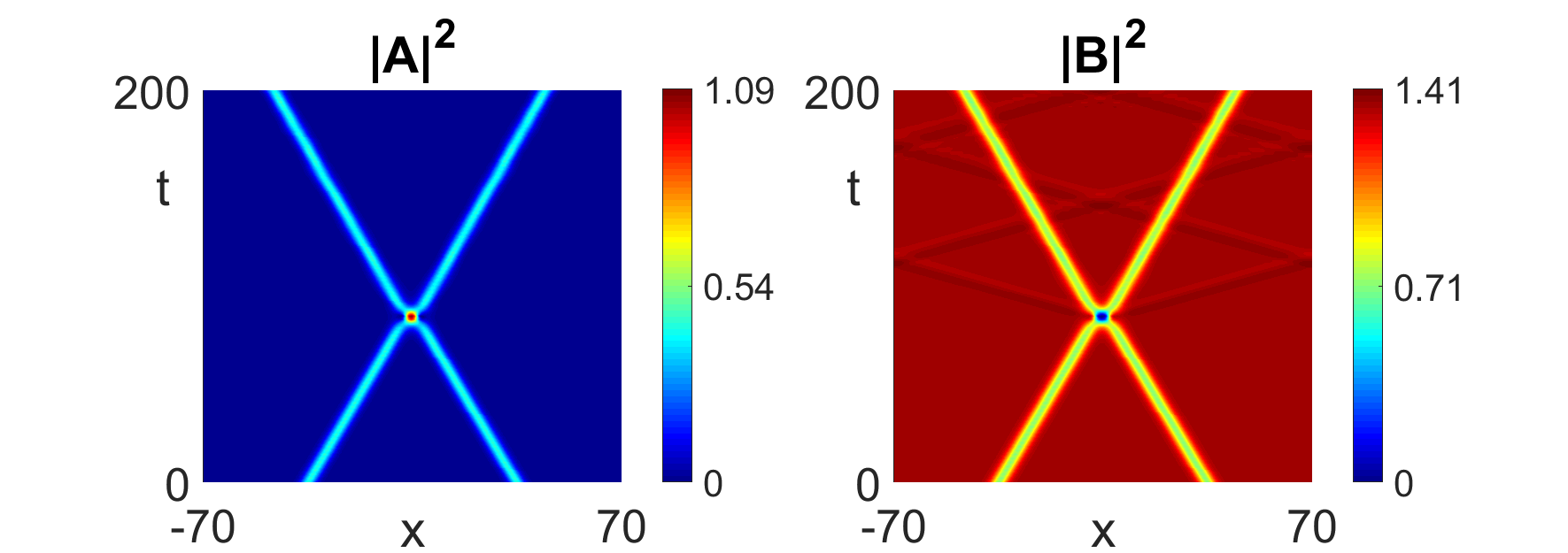}\label{fig:VS_1}}\hfill
  \subfigure[]{\includegraphics[width=.49\textwidth]{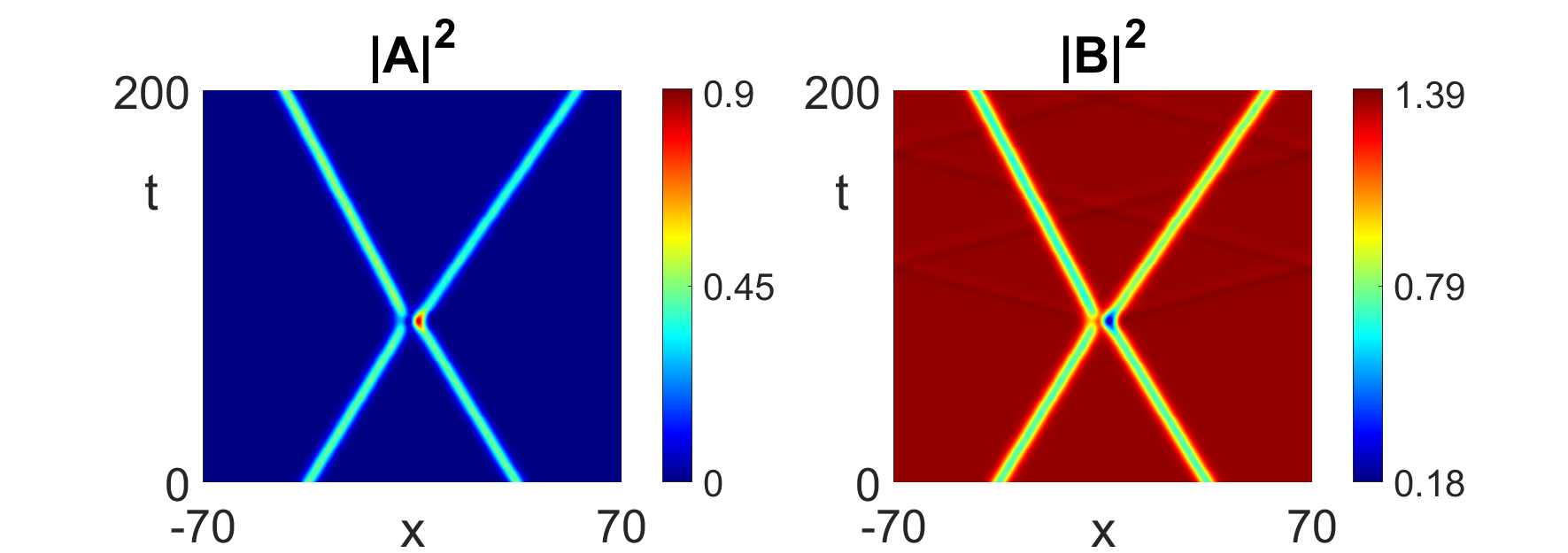}\label{fig:NI_phase_ST}}\hfill
  \subfigure[]{\includegraphics[width=.49\textwidth]{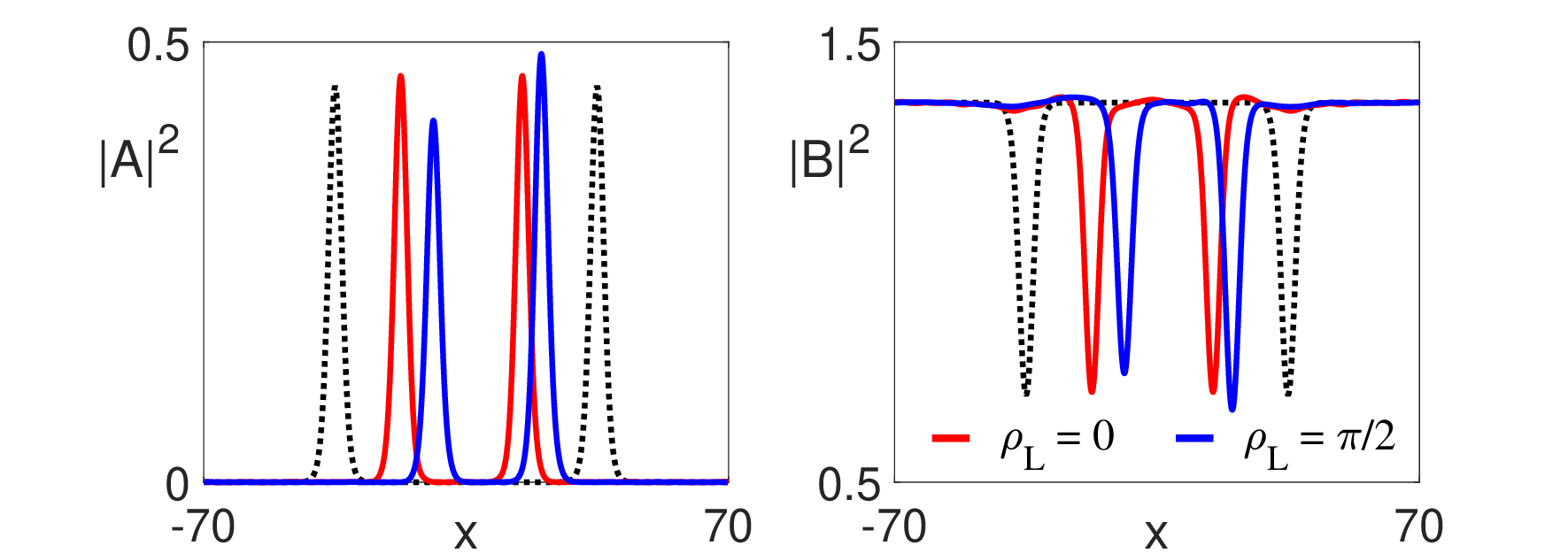}\label{fig:NI_phase_profiles}}
    \caption{(Color online) Collisions between two complex DB solitons with different polarizations. The CNLS parameters are $d_1=d_2=-1$, $g_1=g_2=g_3=2$, and $g_4=3$. The traveling wave parameters $(\omega_A, \omega_B, k_B)=(-9.7,-10,-2.697)$ are the same for both solitons, the velocities of the left and right solitons are respectively $(C_g^{(1)},C_g^{(2)})=(5.789,5)$, and the polarization of the right soliton is $\rho_R=0$. (a, b) Space-time plots in the frame comoving with velocity $C_g^{(0)}=5.395$. The polarization of the left soliton is $\rho_L=0$ in panel (a) and $\rho_L=\pi/2$ in panel (b). (c) Soliton profiles after the unpolarized collision in panel (a) (red), after the polarized collision in panel (b) (blue), and before collisions (black-dashed). The left soliton is always shown in the left half-plane, $x<0$, and the right soliton is always shown in the right half-plane, $x>0$.}
    \label{fig:BD_phase_collision}
\end{figure}

The initial conditions to be time-evolved in Eq.~(\ref{Eq:GCNLS}) consist of these two complex DB solitons respectively located on the left and on the right. The left bright component is multiplied by $e^{i\rho_L}$, and the right bright component is multiplied by $e^{i\rho_R}$, where $\rho_{L,R}\in [0,2\pi)$ are the polarizations of the two colliding solitons. We can set $\rho_R=0$ without loss of generality and explore the effect of varying $\rho_L$. The collision dynamics for two choices of $\rho_L$ are shown in Fig.~\ref{fig:BD_phase_collision}. As expected, the $\rho_L=0$ case yields no mass exchange as shown in Fig.~\ref{fig:VS_1}. To explore $\rho_L\neq0$, we choose $\rho_L=\pi/2$, and observe a mass exchange as shown in Fig.~\ref{fig:NI_phase_ST}. Note that both collisions produce negligible radiation compared with the solitons, which is consistent with the near integrability of Eq.~(\ref{Eq:GCNLS}).

We shall examine this mass exchange in terms of both soliton parameters and soliton profiles.

In terms of soliton parameters, the 2 parameters of the dark component, $(\omega_B,k_B)$, cannot change since the incoming and outgoing solitons must share the same plane wave background. However, the soliton velocity $C_g$, which is a shared parameter between both components, can change as shown in Fig.~\ref{fig:NI_phase_ST}. The 2 parameters of the bright component, $(\omega_A,\rho$), can also change.

In terms of soliton profiles, the profiles of both components can change. As shown in Fig.~\ref{fig:NI_phase_profiles}, there is a visible difference in the amplitudes of both components after the polarized collision in Fig.~\ref{fig:NI_phase_ST} (blue) compared with either after the unpolarized collision in Fig.~\ref{fig:VS_1} (red) or before collisions (black-dashed). Thus, in terms of soliton profiles alone, it is tempting to describe the mass exchange as an energy transfer, but the latter usually applies between two bright components, not between a bright component and a dark one.

Overall, the change in soliton profiles in both components results from the change in the 2 soliton parameters $(C_g,\omega_A)$, as exemplified by Fig.~\ref{fig:bif_diagrams}. The polarizations $\rho_{L,R}$ of the two solitons can also change, but they do not affect the soliton profiles. Thus, the mass exchange is better described by the change in soliton parameters than the change in soliton profiles, although the soliton profiles are more visible than the soliton parameters.

We stress that Fig.~\ref{fig:NI_phase_ST} shows a collision between two generic DB solitons whose dark and bright components have comparable amplitudes. In such cases, the relative changes of the amplitudes of both components produced by the mass exchange are moderate as shown in Fig.~\ref{fig:NI_phase_profiles}. However, if either or both of the two colliding solitons are near bifurcation points, then their dark and bright components could have very different amplitudes. In such cases, the relative changes of the amplitudes of either or both components produced by the mass exchange could be very large or very small. The eigenfunctions in the spectra of vector solitons can yield valuable insights into the collision dynamics of these solitons in non-integrable regimes of Eq.~(\ref{Eq:GCNLS}); see e.g.~Ref.~\cite{yang2000fractal} for the collision dynamics of BB solitons explained from this perspective.

% The existence of energy transfer between vector solitons in the non-integrable CNLS system may provide new opportunities and insight into the possibility of collision-based computing \cite{rand2012computing}, however, useful collision-based computation must entail restoration of full-energy solitons \cite{jakubowski2002computing}.

\section{Conclusion}\label{sec:conclusion}

In this paper, we have outlined a program to classify DWs and vector solitons in the 1D two-component CNLS equation without restricting the signs or magnitudes of any coefficients. The CNLS equation is reduced first to a complex ODE, and then to a real ODE after imposing a restriction. In the real ODE, we have identified four possible equilibria including ZZ, ZN, NZ, and NN, and analyzed their spatial stability. We have identified two types of DWs including asymmetric DWs between ZZ and NN and symmetric DWs between ZN and NZ. We have identified three mechanisms for generating vector solitons in the real ODE including heteroclinic cycles formed by assembling two DWs back-to-back, local bifurcations on the four equilibria, and exact solutions that are scalar solitons with vector amplitudes. Local bifurcations are further divided into the Turing bifurcation that generate Turing solitons and the pitchfork bifurcation that generates DB, BAD, DD, and DAD solitons. We have identified a 3D parameter space for real vector solitons and a 5D parameter space for complex vector solitons, and introduced a numerical continuation method to find these vector solitons. Our classification program applies not only to optics and BECs, but also to the general problem of two interacting quasi-monochromatic plane waves. Further solutions that can be incorporated into our classification program include those originating from secondary bifurcations or codimension-2 bifurcations.

As an application of our classification program, we have shown that real and complex DB solitons can be parametrically related to DWs in the immiscible regime of the CNLS equation relevant to BECs. We have also shown that in this regime, collisions between two polarized DB solitons typically feature a mass exchange that changes the frequencies and polarizations of the two bright components and the two soliton velocities.

Our classification program can be generalized to the $N$-components CNLS equation for $N$ interacting quasi-monochromatic plane waves. This equation overlaps with the $N$-component GPE for bosonic BECs \cite{kevrekidis2016solitons} in certain parameter regimes. This equation has $2^N$ possible equilibria, so its complexity increases exponentially with $N$.

Our classification program is also relevant to the 2D two-component CNLS equation. In the BEC setting, the 2D GPE can exhibit radial solutions like vortices due to the radial symmetry of the Laplacian \cite{kevrekidis2016solitons}. However, for two interacting quasi-monochromatic plane waves in 2D, the dispersion relation of either plane wave can be either elliptic or hyperbolic, so the linear part of the 2D CNLS equation generally lacks radial symmetry. Nonetheless, the 2D CNLS equation can be reduced along any line to the 1D CNLS equation that we studied, so 1D solitons in the latter correspond to 2D line solitons in the former.

We have applied our classification program to DWs and DB solitons in the homogeneous GPE, but the effects of a trapping potential on such solutions can also be remarkable. Stable pinning of DWs near maxima of the potential is shown both physically \cite{dror2011domain} and mathematically \cite{alama2015domain}. A plethora of DB solitons are found with a harmonic trap using numerical continuation in recent studies \cite{wang2021dark,wang2022systematic}. The GPE in the immiscible regime with a trapping potential continues to provide a great venue for further explorations of DWs and DB solitons.

\section*{Acknowledgement}

YM acknowledges support from a Vice Chancellor's Research Fellowship at Northumbria University. We thank Gennady El, Benoit Huard, Antonio Moro, and Matteo Sommacal for their helpful comments.

\section*{References}
\newcommand{\newblock}{}
\bibliographystyle{unsrt}
\bibliography{references.bib}

\end{document}